\documentclass[sn-mathphys,iicol]{sn-jnl}

\usepackage{float}
\usepackage{subfigure}
\usepackage[draft]{minted} 
\jyear{2022}%

\theoremstyle{thmstyleone}%
\theoremstyle{thmstyletwo}%

\theoremstyle{thmstylethree}%

\begin{document}

\title[Nebula Graph]{Nebula Graph: An open source distributed graph database}


\author*[1]{\fnm{Min} \sur{Wu}}\email{wumin@hdu.edu.cn}
\author[2]{\fnm{Xinglu} \sur{Yi}}\email{yee.yi@vesoft.com}
\author[2]{\fnm{Hui} \sur{Yu}}\email{jerry.yu@vesoft.com}
\author[3]{\fnm{Yu} \sur{Liu}}\email{lionel.liu@vesoft.com}
\author[3]{\fnm{Yujue} \sur{Wang}}\email{lionel.liu@vesoft.com}


\abstract{
This paper introduces the recent work of Nebula Graph, an open-source, distributed, scalable, and native graph database. We present a system design trade-off and a comprehensive overview of Nebula Graph internals, including graph data models, partitioning strategies, secondary indexes, optimizer rules, storage-side transactions, graph query languages, observability, graph processing frameworks, and visualization tool-kits. In addition, 
three sets of large-scale graph benchmark tests are conducted.
}




\maketitle

\section{Introduction}


In the last decade, new technological applications and development such as social media networks, banking fraud and money laundering networks, coronavirus contact tracing, and knowledge graphs have generated a huge amount of highly-related data. Traditional relational databases (RDBMS) cannot process such in-depth complexity for real-time responses.
Thus, graph databases (GDBMS), wherein entities of interest are represented by \textit{vertices} or \textit{nodes} and relationships between them by \textit{edges} have recently regained high prevalence 
\cite{dbengineranking}. Furthermore, as the amount of data produced is enormous and ever-increasing, distributed graph databases are inevitable to address problems of a big amount of data, rather than a single server solution. 

This paper introduces the recent work of Nebula Graph (v3.0), an open-source\footnote{Apache License, Version 2.0.}, distributed, scalable, and native graph database. It is capable of hosting graphs with hundreds of billions of vertices and trillions of edges, and serving queries with millisecond-latency. It is capable of scaling both vertically and horizontally, and automatically sharding a large graph into multiple subsets stored in multiple servers. It also has high availability backed by the Raft protocol\cite{ongaro2015Raft}. 

\section{Nebula Graph Architecture}

Nebula Graph consists of three major components: \textit{Nebula Core}, \textit{Nebula Algorithms \& Analytics}, and \textit{Clients, Tools \& Visualization}. In this paper, we mainly focus on the internals of Nebula Core, with some explanations of the other two components.

\begin{figure}[ht]
  \centering
  \includegraphics[width=\linewidth]{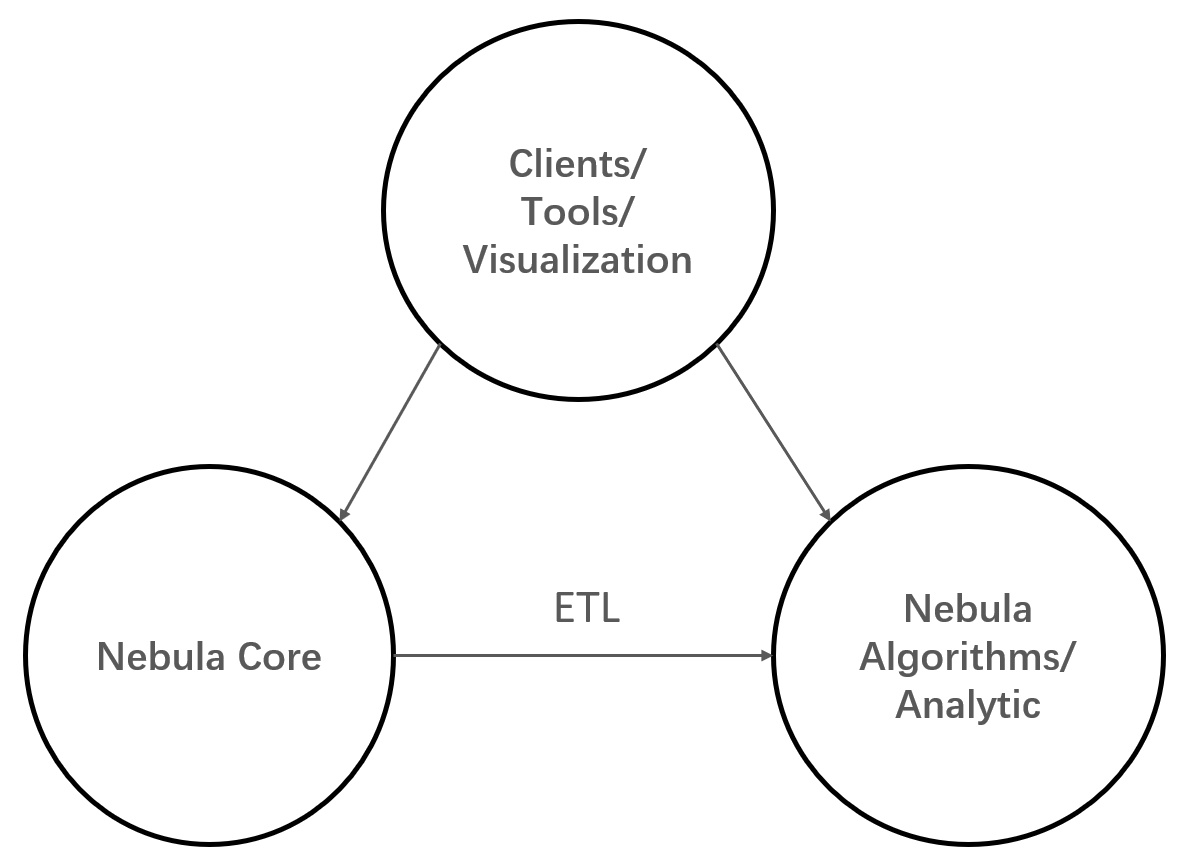}
  \caption{Major components of Nebula Graph.}
  \label{fig:components}
\end{figure}

\subsection{Nebula Core} \label{Sec:Core}

Nebula Core serves as the OLTP (online transaction processing) part of the Nebula Graph, which is responsible for answering enormous queries within a local scope of the graph, such as neighborhoods, certain restricted traversals, paths, sub-graphs, lookups, inserts, deletes, and updates of a set of vertices and edges. The goal of Nebula Core, which is written in C++, is to achieve high throughput (TPS/QPS) with scalability as the top design priority.

To achieve the best performance in a large cluster, Nebula Core applies a separation architecture for storage and compute. It consists of three services: the \textit{Meta Service}, the \textit{Storage Service}, and the \textit{Graph Service}. 

Each service has its own executable binaries that can be hosted on a Kubernetes cluster\cite{k8s}. Therefore, users can deploy a Nebula Core cluster on a single server or multiple servers.

\begin{figure}[ht]
  \centering
  \includegraphics[width=\linewidth]{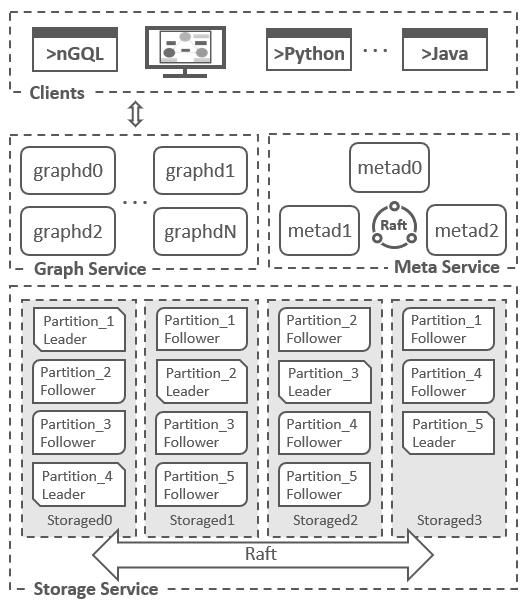}
  \caption{The Architecture of Nebula Core.}
  \label{fig:arch}
\end{figure}

Fig.~\ref{fig:arch} shows the architecture of a typical Nebula Core.

The Meta Service is responsible for metadata management, such as schema operations, cluster administration, user privileges, and license management.

All the meta processes form a \textit{Raft-protocol} group, which initiates an election with quorum-based periodical terms. In the group, there are some roles, (Raft-)\textit{leader}, \textit{followers},  \textit{listeners}, and \textit{drainers}. The leader is unique. It is elected by all of the candidates: the leader and followers. Listeners and drainers do not participate in the election. Only the leader can provide services to the clients or other components of Nebula Graph. The followers run in a standby way and each has the data replication of the leader. Once the leader fails, usually due to a heartbeat timeout, one of the followers will be elected as the new leader. The listeners and drainers will be explained in Sec.~\ref{sec:index} and \ref{sec:datacenter}.

The Storage Service stores all the graph data. There are three layers: the \textit{storage interface}, which defines a set of APIs that are related to the graph operations (i.e., get neighbors, lookup, insert, delete, et al.), the \textit{consensus layer}, which implements the Raft protocol to ensure the strong consistency between data replicas and high availability, and a \textit{key-value library}, backed by RocksDB\cite{rocksdb} for now.

The Graph Service is stateless. It supports two SQL-like declarative languages, openCypher\cite{cypher} (DQL) and nGQL (a Nebula-native query language, which implements DML, DCL, etc.).
A query is executed by a single process with multiple threads. But more graph processes can join to serve more queries. The workflow of the Graph Service is discussed in Sec.~\Ref{sec:opt}.

\subsection{Data Model}

Nebula Graph follows the most popular graph data models in industry, \textit{property graph}, with some extensions and variants. 

\begin{itemize}
\item {\verb|Graph spaces|}: A graph space is used to isolate data from different teams or projects. It defines property data types, storage replications, privileges, data partitions, etc. 
\item {\verb|Vertex|}: Vertices are identified with a graph-space-scope unique identifier, a.k.a. the VID (or Vertex ID). VIDs are provided by users.
\item {\verb|Tag|}: Tags are used to categorize vertices. A tag defines a set of given property types. A vertex can have zero or more tags. It is similar to \textit{labels with constrains} in Neo4j\cite{neo4jsite}.  
\item {\verb|Edges|}: An edge is a connection between two vertices. An edge must have a direction.
\item {\verb|Edges Types|}: Multigraphs are also supported (Multiple edges between two vertices are allowed). The edge type defines a given set of property types, and gives a name to this type of edges. An edge must have one and only one edge type.
\item {\verb|Edges Rank|}: The rank value is an immutable user-assigned 64-bit signed integer. It distinguishes the edges with the same edge type between two given vertices. An edge must have one and only one edge rank.
\item {\verb|Property|}: A property is a key-value pair. Properties are defined by tags or edge types. SQL-like data types, such as numeric, boolean, string, text, date-time, NULL, and geo-spatial are supported. 
\end{itemize}

\subsection{Graph Partitioning and Data Format} \label{sec:partition}

Since a very large graph cannot be fully stored and processed on a single server, graph partitioning is inevitable. Though there is plenty of research and algorithms on the optimal (e.g, min-cut) graph partitioning, streaming/dynamic graph partitions are few. Furthermore, graph partition algorithms are well known to be NP-hard and seldom used in industrial OLTP graph databases. 

Therefore, Nebula Core chooses a straightforward way to model vertices and edges as key-value pairs. So it can rely on a lot of features of sophisticated distributed key-value systems. 

And the number of partitions is determined once a graph space is created. 
Every server can host multiple partitions (usually 2-20), and the partitions can migrate between servers for data balancing and workload balancing. But key-value pairs can not migrate between partitions since they are located by a\textit{ static VID hashing} strategy.

\begin{figure}[ht]
\centering
\includegraphics[width=\linewidth]{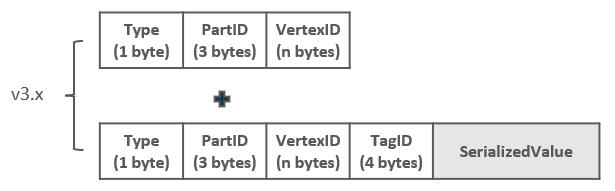}
\caption{Vertex key format}
\label{fig:vertex}
\end{figure}

Specifically, a vertex is modeled as key-value pairs (Fig.~\ref{fig:vertex}): One key (\textrm{Type+PartID+VertexID}), which represents the vertex itself, denotes the vertex's partition ID and its VID, and the value is empty. Optionally, in addition to the same prefix, the other key (\textrm{Type+PartID+VertexID+TagID}), which represents the properties of a specific tag, includes a TagID which denotes the tag of the vertex. All the properties defined by this TagID are serialized as the value part. If this vertex has one more tag (and therefore properties), then one more key and the serialized value are added accordingly. See also Table.~\Ref{table:vertex}.

\begin{figure}[ht]
\centering
\includegraphics[width=\linewidth]{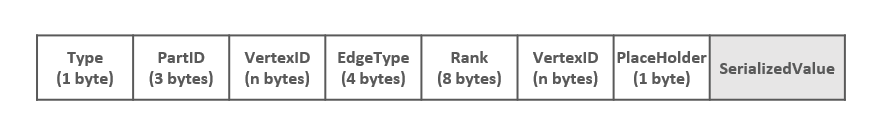}
\caption{Edge key format}
\label{fig:edge}
\end{figure}

\begin{figure}[ht]
\centering
\includegraphics[width=\linewidth]{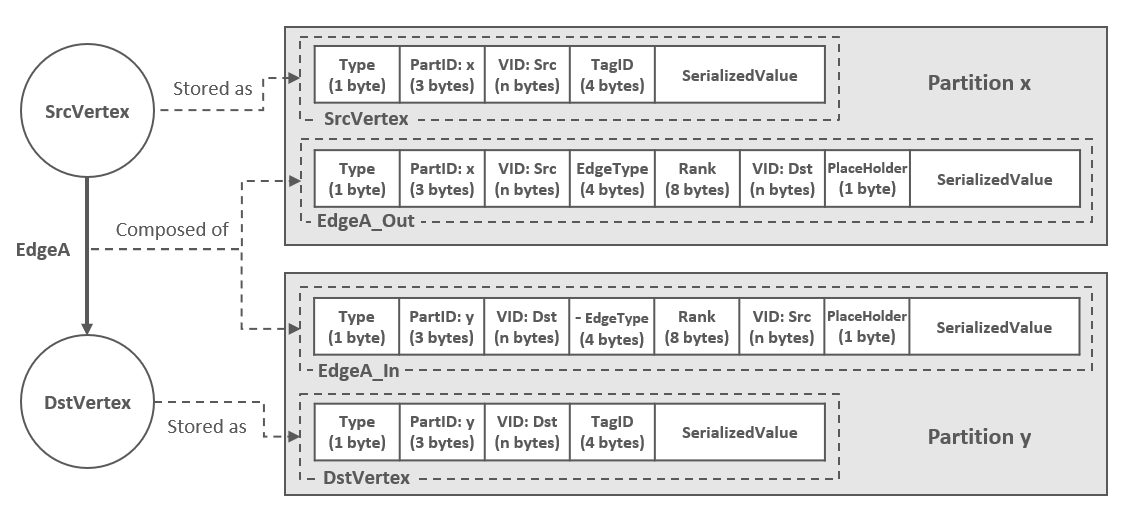}
\caption{An illustrative example of a source vertex, a destination vertex, and a directed edge, with the corresponding key-value pairs on two partitions. They are partitioned by a static VID hashing strategy.}
\label{fig:edge-division}
\end{figure}

In the format of an edge shown in Fig.~\Ref{fig:edge}, the PartID is the same as that of the source vertex, and is followed by the Vertex's VID, EdgeType, Edge-Rank, destination vertex's VID, and a reserved placeholder for MVCC. The properties are also serialized in the value part. Besides, to facilitate reverse traversing, from a destination vertex back to a source vertex, a reversing edge is automatically inserted. The positions of source and destination VIDs are swapped. See Fig~\Ref{fig:edge-division}.

It's worth noting that a vertex and its neighboring edge keys share the same prefix. So in most key-value systems, it is equivalent to a partitioned \textit{adjacency list}. Therefore the storage and processing locality between multiple servers can be well guaranteed. It also works for most breadth-first search algorithms, which can be concurrently executed as the traversal depth increases.

Moreover, the \textit{graph structure} can be entirely represented by the edge's key part, without the need of the value part, which represents edge properties. The graph-structure-based traversing is the most important query pattern\cite{gql2017} in the sense of a graph database. Therefore, some key-value separation technologies\cite{lu2017wisckey} can be introduced to accelerate read latency. All the keys (both the vertex and edge keys) can be cached in the memory to gain better performance, and the serialized values are maintained in the SSD disks (and partially cached in main memory). See the benchmarking in Sec.~\Ref{sec:bench}.


\subsection{Secondary Indexing on Property} \label{sec:index}

Nebula Graph is designed to be a primary GDBMS for applications. Vertex keys and edge keys are primarily indexed as a classic log-structured-merge (LSM) tree by key-value libraries. The properties can either be directly fetched through VIDs (key), or indirectly accessed through the secondary indices. 

There are three types of secondary indices in Nebula Graph: \textit{native index}, \textit{full-text index}, and  \textit{geography index}. The former index relies on the same key-value library (RocksDB) as explained in Sec.~\Ref{Sec:Core}. And the full-text index relies on a Raft-listener with Elastic Search\cite{gormley2015elasticsearch}. Geo-spatial indexes use Google's S2 library\cite{googles2}, which is straightforward and will not be illustrated in this paper. But we'd like to mention that unique indexes or constraints are not implemented in Nebula Graph.

Basic property data types (i.e., numeric, boolean, string, date-time) can be accessed by native indexes. An illustrative example of vertices and their indices are given in Table.~\Ref{table:vertex} and \Ref{table:index}. We insert two example vertices (VID 50, 60) with two Tags. The example of the corresponding key values are shown in Table.~\Ref{table:vertex}. 

\begin{minted}[frame=lines, framesep=2mm]{SQL}
-- tag-a defines three properties: pa-1,
-- pa-2, pa-3. and example inserted value
-- is ta-1, ta-2, ta-3;
INSERT VERTEX tag-a (pa-1, pa-2, pa-3) 
VALUES 50: (ta-1, ta-2, ta-3)

-- insert two vertices 50 and 60 with the 
-- same tag-b, which defines two properties:
-- pb-1, pb-2, and the same value: tb-1, tb-2
INSERT VERTEX tag-b (pb-1, pb-2) VALUES 50:
(tb-1, tb-2)
INSERT VERTEX tag-b (pb-1, pb-2) VALUES 60:
(tb-1, tb-2)

-- create a composite index on pa-1 and pa-2
-- but not on pa-3.
CREATE TAG INDEX i-a ON tag-a (pa-1, pa-2);
-- create a single index on pb-2
CREATE TAG INDEX i-b ON tag-b (pb-2);
\end{minted}

\begin{table}[ht]
\centering
\caption{The example vertices}
\begin{tabular}{|lll|l|}
\hline
\multicolumn{3}{|l|}{Key} & Value               \\ \hline
PartId & VertexId & TagId & Serialized property \\ \hline
100    & 50       &       &                     \\ \hline
100    & 50       & tag-a & ta-1 ta-2 ta-3      \\ \hline
100    & 50       & tag-b & tb-1 tb-2           \\ \hline
101    & 60       &       &                     \\ \hline
101    & 60       & tag-b & tb-1 tb-2           \\ \hline
\end{tabular}
\label{table:vertex}
\end{table}

\begin{table*}[hbt]
\centering
\caption{The vertices' indices}
\begin{tabular}{|lllll|l|}
\hline
\multicolumn{5}{|c|}{Key}                                                                                                                                            & Value \\ \hline
PartId & Index-Id        & Index-Binary             & Index-Length                  & VertexId &       \\ \hline
100    & i-a            & ta-1 ta-2 & length(ta-1) length(ta-2) & 50       &       \\ \hline
100    & i-b            & tb-2      & length(tb-2)              & 50       &       \\ \hline
101    & i-b            & tb-2      & length(tb-2)              & 60       &       \\ \hline
\end{tabular}
\label{table:index}
\end{table*}

The format of index is shown in Table \Ref{table:index}. The vertex 50 and all its indices are stored in the same partition for locality.  Index-Id tells the related tags and properties. Index-Binary is the serialized binary format of corresponding properties. The Index-Length can segment the Index-Binary by each property's length (E.g., "ab"+"cd" or "abc" + “d”). The VertexId is not put in the value part but in the key part. Because two vertices may have the same tag, the same properties and in the same partition by chance. Therefore, the vertex's index is seek-ed by prefix, PartId + Index-Id + Index-Binary + Length, to get the VertexId. The value part of index is empty.

Because the key of the index is unchangeable, to update on the vertex property will trigger a long sequence of I/O operations: read old index, delete old index, write new index, and write/update vertex. The write performance is slower since the read index step is a random disk operation. So two special operations are supported for massive initial loading of the graph space: \textrm{IGNORE\_EXISTED\_INDEX} when inserting vertices without their indexing, and later run \textrm{REBUILD INDEX} to make all indexing updated. It can be seen that both are sequential operations on disks, therefore much faster.

\begin{figure}[ht]
\centering
\includegraphics[width=\linewidth]{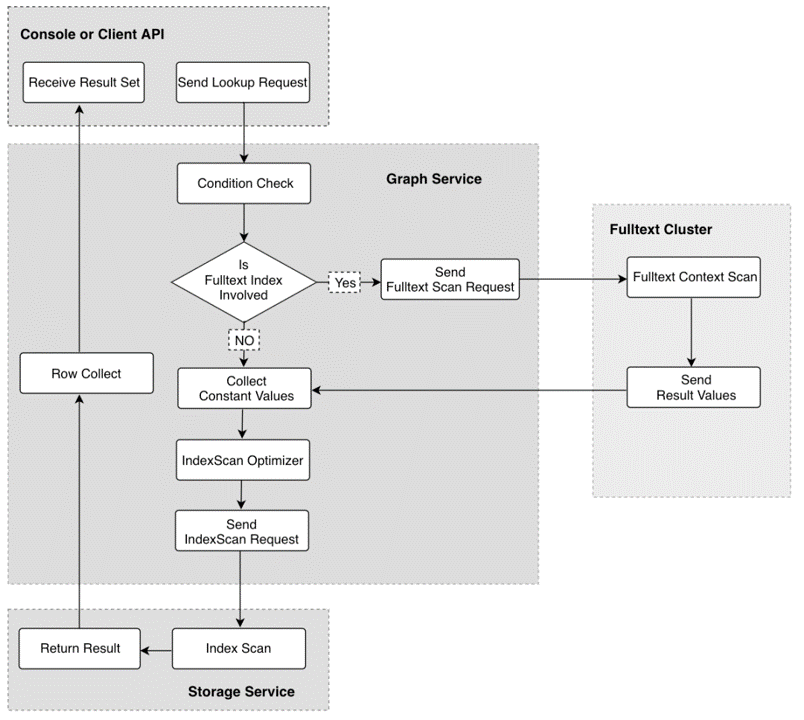}
\caption{Text index redirects to Elastic Search by Graph Service.}
\label{fig:text-index}
\end{figure}

For text-based searches, the native index is not suitable, since text can be much longer and fuzzy search is required. Therefore, a third-party tool Elasticsearch is introduced. To synchronize data from Nebula Core to Elastic Search, a new role called \textit{text-listener} is implemented. In the Raft protocol, a \textrm{listener} works similarly to a Raft follower, but it will not ever engage in the Raft quorum election. The text-listener only subscribes to the write-ahead log (WAL) from the leader and asynchronously sends data to an Elasticsearch cluster. And the text-based index searches will be redirected to the Elasticsearch cluster by Graph Service as in Fig.~\Ref{fig:text-index}. 

For a read request, the index-choosing strategy is implemented in the Optimizer (Sec.~\ref{sec:opt}) in Graph Service. And it is a set of rules: single property/single index comparison (=, $<$, etc), single property prefix comparison (equal, less, start from, etc), multi-property/composite index comparison (composite index), text fuzzy comparison. And if no index is available, a full vertex/edge scan will be conducted. The matching rule is chosen in the above order. More details about these optimization rules will be discussed in Sec.~\ref{sec:opt}.

\subsection{Computation Pushing Down and Optimizer Rules} \label{sec:opt}

\begin{figure}[ht]
\centering
\includegraphics[width=\linewidth]{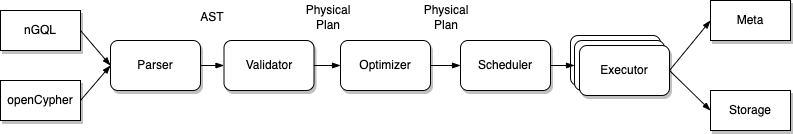}
\caption{Workflow of Graph Service}
\label{fig:GraphEngine}
\end{figure}

In the early version of Nebula Graph (v1.0), all the nGQL clauses can be directly mapped to physical executors. This is hard for the Graph Service to rewrite and generate an optimized execution plan especially when a query is long and complex (e.g. LDBC-SNB tests\cite{LDBC-SNB}). So the performance heavily depends on how the user writes the query sentences. Furthermore, as being compatible with openCypher in v2.0, these two languages can generate different plans with different executors. These make the system too complex and hard to profile queries. Therefore, the classical "\textit{parser,optimizer,executor}" workflow in DBMS have been introduced in Nebula Graph v2.0 (Fig.~\Ref{fig:GraphEngine}). These two front-end languages can share the same logical optimizer rules to generate a unified execution framework (back-end). 

When a query statement is sent to the Graph Service, it is parsed and validated to generate an abstract syntax tree, AST, and then the initial plan. The initial plan is optimized by the \textit{Optimizer} to be physical plans, which are executed by a group of executors. Except for the basic executors as in relational algebra (e.g., selection, projection, Cartesian product, union, and difference), additional graph-aware executors (totally 113) are introduced\footnote{https://github.com/vesoft-inc/nebula/tree/release-3.1/src/graph/planner/plan}. For example, \textit{GetNeighbors} gets the neighboring VIDs of a given VID, \textit{GetProps} gets the property of a vertex or an edge, and \textit{Loop} matches the variable-length path. 

Because of Nebula Graph's separate architecture for storage and compute, minimizing the remote data transfer from Storage and Graph Services is a major performance issue to be considered in the design of optimizer. The other goal is to search and choose a optimized sequence of executors to replace initial less-efficient executors according to predefined rules. That is a rule-based optimizer (RBO).

\begin{figure}[ht]
\centering
\includegraphics[width=\linewidth]{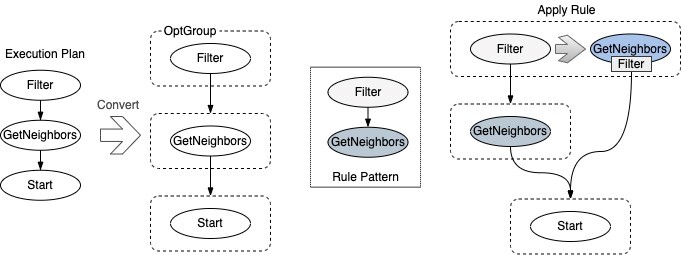}
\caption{An example: filer-push down rule is applied to generate a new sub-plan.}
\label{fig:optimizer}
\end{figure}

The Graph Service implements the cascades framework\cite{graefe1995cascades} for query optimization. In this framework, there are two steps: \textit{exploration} and \textit{transformation}. In the previous step, all rules are applied to the initial plan and expanded to a larger forest of plan trees. To implement this step, two utility classes are designed, \textit{OptGroup} and \textit{OptGroupNode}:
An OptGroup contains a group of OptGroupNodes. In the same OptGroup group, all OpGraphNodes are considered to be equivalently exchangeable. Therefore the Optimizer can choose a better OptGroupNode. Every OptGroupNode depends on other OptGroups, but not OptGroupNode. The arrows in Fig.~\ref{fig:optimizer} show the dependency. All the dependencies form a plan forest. 

In the exploration step, from the leaves of this forest, every pattern rule is applied to all the fragments (sub-plan) of the forest. If a sub-plan matches this pattern, this sub-plan is replaced to a new sub-plan. Both the match and replacement are predefined in the optimization rules. This new sub-plan is considered to be better than the original sub-plan. Every rule defines three interfaces:

\begin{minted}[frame=lines, framesep=2mm]{C++}
struct OptRule {
  // This plan Node matches
  // the rule's pattern
  bool match(OptGroupNode);
  
  // Transform current sub-plan
  // fragments to generate a new one
  Result transform(MatchedResult);
  
  // Current rule pattern definition
  Pattern pattern();
}
\end{minted}
After all rules are applied in the exploration step, the optimized physical plan is considered to be generated.

The comprehensive rules are listed here\footnote{ https://github.com/vesoft-inc/nebula/tree/release-3.1/src/graph/optimizer/rule}.
Furthermore, to implement the popular Cost-Based Optimizer (CBO), some preliminary statistics should be collected or sampled, including hardware (memory, CPU, network traffic, and I/O. See Sec.~\Ref{sec:security}), metadata (in-degree, out-degree, distribution, index cardinality, etc). The latter is still working on and not available currently. 

\subsection{Transaction on Storage Side (TOSS)}

Nebula Graph chose to support the style of NoSQL databases. Therefore the ACID in sense of DBMS is not full-filled due to some write/read performance considerations of a distributed system. Especially, the isolation level is not guaranteed. But as shown in Fig.~\Ref{fig:edge-division}, a single insertion of one edge will be amplified to two key-value (out-edge and in-edge) writes. To guarantee the atomic remote operation of these two key-value pairs, a light-weight technique inspired by\cite{escriva2015warp} and two-phase locking (2PL) called \textit{Transaction on Storage Side} (TOSS) is introduced.

Usually, two partitions are engaged in the writing procedure. One partition (out-edge partition) serves the out-edge key-value pair, and the other (in-edge partition) serves the in-edge key-value pair. Besides, every partition can have multiple replicas (Raft leader and followers) based on the Raft protocol for avoiding possible crashes. These two partition leaders are usually located on two different servers.

\begin{figure}[htb]
\centering
\includegraphics[width=\linewidth]{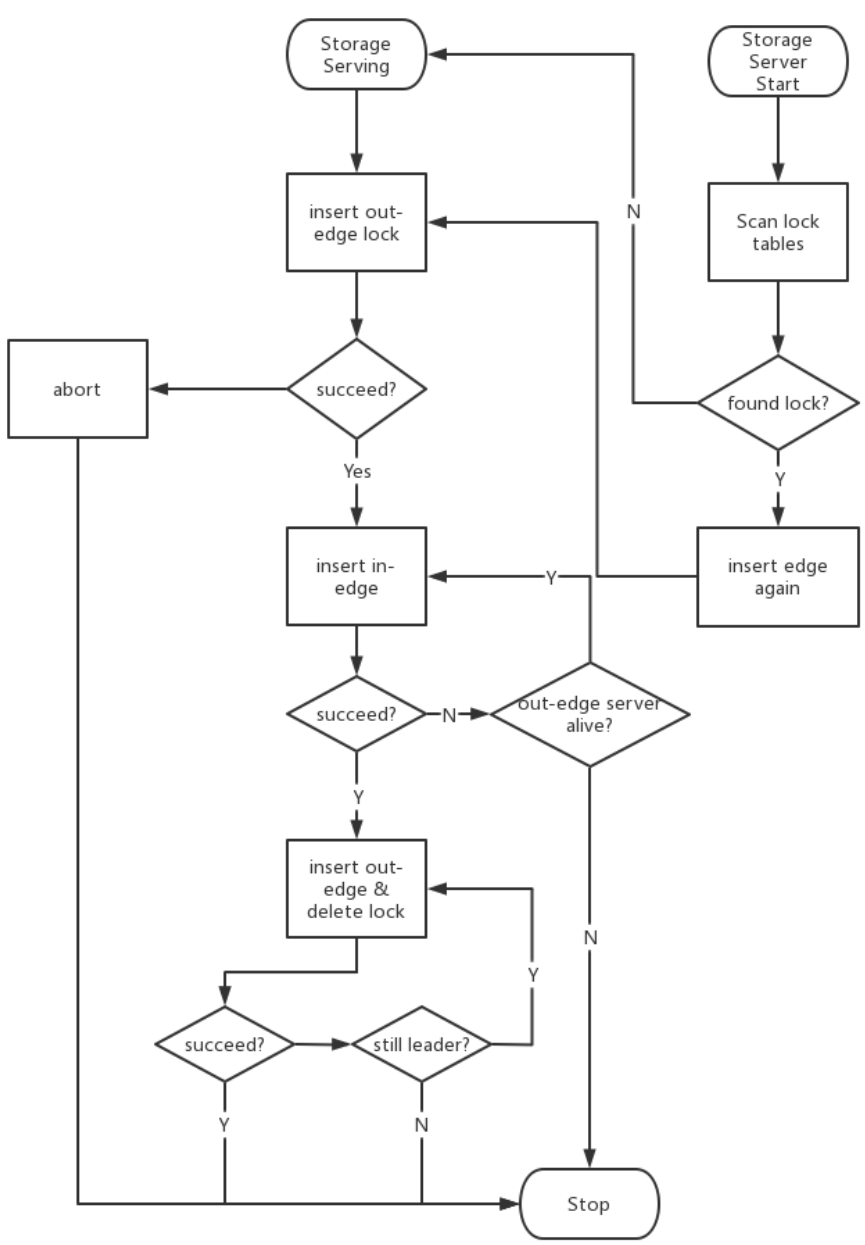}
\caption{Workflow of TOSS.}
\label{fig:toss}
\end{figure}

The out-edge partition receives an edge write request from the client, and then it starts a spinning lock (\textrm{SRC\_VID+EdgeType+Rank}) for itself. Then a message is sent by the out-edge partition to the in-edge partition to ask for writing the in-edge key-value pair. 
After the write's success callback from the latter is acknowledged, the out-edge partition unlocks.
But there could be possible failures or crashes during this procedure.
Therefore the Raft protocol is used as well. The out-edge partition's leader commits a special item for this lock to all the followers when locking/unlocking. Even if any crash occurs, all engaged roles can replay the states according to Raft and continue the procedure until a success. Unlike DMBS transaction, no rollback takes place in this procedure. The workflow is shown in Fig.~\Ref{fig:toss}.

\subsection{High Availability Crossing Data Centers}  \label{sec:datacenter}

\begin{figure*}[thb]
\centering
\includegraphics[width=\linewidth]{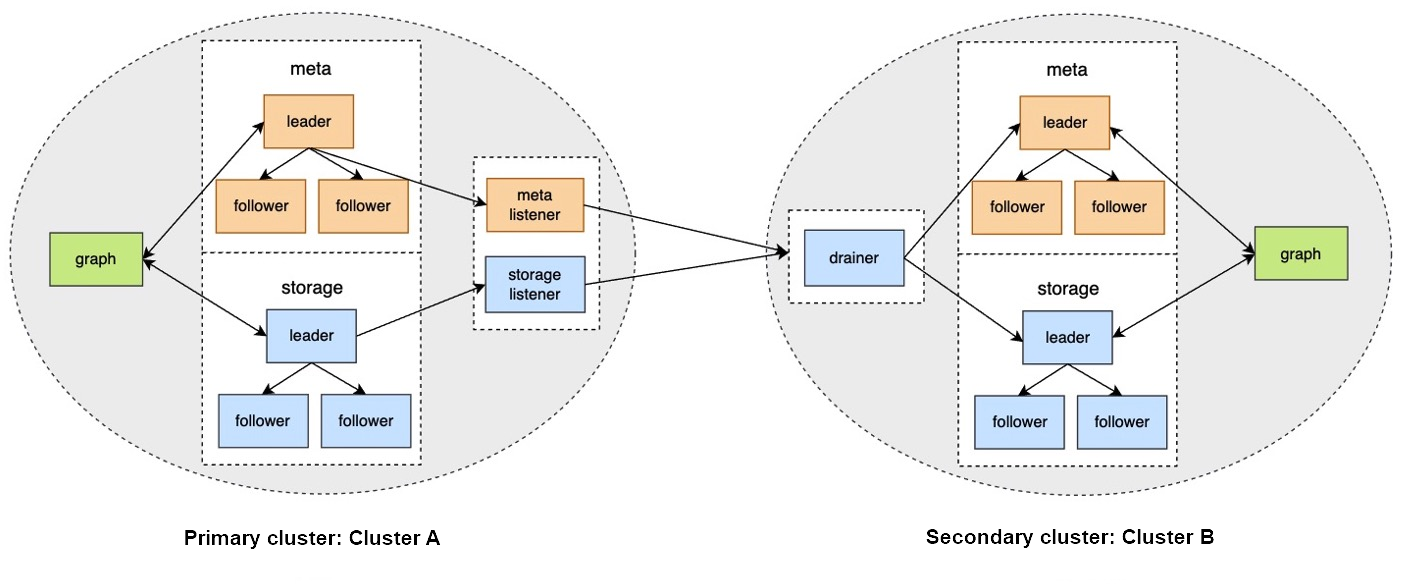}
\caption{Data flows between two clusters.}
\label{fig:twocenters}
\end{figure*}

Usually, one cluster of Nebula Graph should be deployed inside a data center due to the network fluctuation and heartbeat timeout constraints. And to populate persistent meta and data crossing multiple data centers, three roles, a \textit{data-listener}, a \textit{meta-listener} and a \textit{drainer} are implemented (Fig.~\Ref{fig:twocenters}). 

As an example, a meta listener is deployed to subscribe to the Raft-WAL from Meta Service, and a storage listener for the data from the Storage Service. Usually, these two types of listeners are deployed together with the primary cluster (the major cluster for the application's writing). And then these two listeners write asynchronously to a remote drainer who is deployed in the same data center with the secondary cluster (the read-only cluster). Basically, the drainer works as a client of the secondary cluster. So the two clusters are working independently, and the configurations and deployments of them both can be heterogeneous. 

Both listeners and drainers are stateful, they all have their own WAL for possible failures. In order to deploy more read-only secondary clusters, a chained-populate method is suggested, i.e., 1-to-1 cluster synchronization. There are some registration operations on every role for the upstream-downstream dependency. Listeners or drainers with the same kind share the overall workload burden by a M-to-N round-robin scheduling (partition-to-listener,
listener-to-drainer).

\subsection{Query Languages}

As mentioned above, Nebula Graph supports two declarative languages, openCypher 9 (DQL part) and nGQL, a Nebula-native query language (v3.0\cite{NebulaManual}). But neither Apache Gremlin\cite{rodriguez2015gremlin} nor W3C SPARQL\cite{world2013sparql} is supported. And there are now two groups (ISO/IEC JTC 1 N 14279 and JTC 1/SC 32 N 3228) working on the international standards of graph query languages, which are planned to be released in 2023. So Nebula Graph is planned to be modified accordingly to meet the upcoming international standards. 

\subsubsection{Data Definition Language (DDL)}

Nebula Graph is strong-typed, and not entirely schema-free as Neo4j. The types and properties of vertices and edges must be defined before any usage. A simplified syntax of creating an edge type (edge's type and its property definition) is shown as follows:
\begin{minted}[frame=lines, framesep=2mm]{SQL}
CREATE EDGE [IF NOT EXISTS]
<edge_type_name> ( <prop_name>
<data_type> [NULL | NOT NULL] )
\end{minted}
E.g., an edge type called \textit{write\_paper} is created with one property (\textit{wtime}): 
\begin{minted}[frame=lines, framesep=2mm]{SQL}
CREATE EDGE write_paper (wtime DATE);
\end{minted}
Except creation (\textit{CREATE EDGE}), statements for dropping (\textit{DROP EDGE}), altering (\textit{ALTER EDGE}), \textit{SHOW EDGES}, and \textit{DESCRIBE EDGE} are also supported. They can be inferred directly from SQL's DDL of tables. The syntax of operations on vertex types (TAG) is similar and neglected.

\subsubsection{Data Manipulation Language (DML)}

After running DDL clauses, data can be written into Nebula Core through DML. And then possible Extract-Transform-Loaded (ETL) to the Nebula Algorithms \& Analytics Framework (Sec.~\Ref{sec:algo}).  A simplified syntax of writing an edge is shown as follows:
\begin{minted}[frame=lines, framesep=2mm]{SQL}
INSERT EDGE <edge_type>(<prop_name_list>) 
VALUES <src_vid> -> <dst_vid>[@<rank>]
: (<prop_value_list>)
\end{minted}
Operations of creation (\textit{INSERT EDGE}) and deletion (\textit{DELETE EDGE}) are implemented. E.g., to insert an edge representing that an author writing a paper,
\begin{minted}[frame=lines, framesep=2mm]{SQL}
INSERT EDGE write_paper (wtime) VALUES 
"author_abc" -> "paper_123"
: (DATE("2022-04-20"));  
\end{minted}
If an insertion's success is acknowledged, the majority of Raft replicas in the Storage Service are written synchronously. But full-text indexes of Elasticsearch and crossing-data-center replications (if possible) are written asynchronously by corresponding listeners. 
If an input parameter does not meet the defined data type (e.g., try to write "hello" as \textit{wtime}), an error will be raised. And the write request fails with nothing happened.

\subsubsection{Security and Data Control Language (DCL)}

Access control of Nebula Graph is role-based. OpenLDAP is implemented for authentication. Clauses \textit{CREATE, DESCRIBE, ALTER, DROP, SHOW on USER} and \textit{GRANT/REVOKE ROLE} are used to control graph-space level privileges of different roles. Except for authentication, data transmission with SSL encryption is supported. 

There are commands to maintain the system. \textit{ADD/REMOVE HOST} is used to scale out/in the cluster; \textit{BALANCE DATA} is used to re-balance data placement according to workload; \textit{CREATE SNAPSHOTS} is used to create a snapshot backup of the database and can be uploaded on cloud (S3) and to restore snapshots by \textit{Backup \& Restore} tools.

\subsubsection{Data types, Functions, and the Data Query Language (DQL)}

Except for basic data types (such as numeric, boolean, string,
text, datetime, NULL, and geo-spatial), some composite data types are also supported (List, Set, Map, Path, and Subgraph). Data types and functions are designed to follow the definition of openCypher. Frequently used (but not all) openCypher DQL clauses (\textit{MATCH}, \textit{OPTIONAL MATCH}, \textit{WITH}, \textit{GROUP BY}, \textit{ORDER BY}, \textit{LIMIT}, \textit{WHERE}) are implemented.

And there are some extensions. Three well-tuned clauses are provided as syntactic sugar: \textit{GO} (breadth-first searches), \textit{FETCH} (get vertex/edge properties), \textit{LOOKUP} (directly use secondary indexes). 

All of these clauses share a common set of physical operators. The compatibility and performance are tested with the Cypher technology compatibility kit (TCK)\cite{openCypherTCK} and LDBC Social Network Benchmark (LDBC-SNB), both of which are the most wildly used test cases in the graph database industry.

\subsection{Nebula Algorithms \& Analytics Framework} \label{sec:algo}

Nebula Algorithms \& Analytics Framework is the OLAP (Online analytical processing) part of Nebula Graph. The OLAP requests are not processed at interactive speed, and the computation is mostly complex and global graph in scope. These computations mostly adopt the typical "Gather-Apply-Scatter" abstraction\cite{gonzalez2012powergraph}. 
 
For now, there are two popular open-source frameworks that are supported natively in Nebula Graph, i.e., Spark/GraphX\cite{gonzalez2014graphx}  and Tencent/Plato\cite{tencentplato}. An ETL progress, which is implemented as a Spark job, is used to transfer subgraph data from Nebula Core directly into these frameworks. The ETL process is aware of the source graph's schema and privileges. It is also aware of the target framework's format (e.g., DataFrame in Spark, or the pre-sorted id's in Plato). Therefore it does not have to store the source graph to an intermediate CSR or CSV file but in the direct data format of the framework.  About 30 graph algorithms like Jaccard similarity, Triangle Count, Node2Vec, etc, are implemented as supplements to the original open source algorithms.

\subsection{Clients, Tools \& Visualization}

Nebula Graph uses FBthrift\cite{watson2014under} as its Remote Procedure Call (RPC) framework. Therefore all available SDKs, including Java, CPP, Python, and Golang must be compiled accordingly. 

Some big data system connecting tools are provided as well, including the connectors for Apache Spark, Flink, and Kafka. Users can choose two built-in tools, Nebula Importer and Nebula Exchange to import/export multiple data sources to/from Nebula Graph, including CSV, JSON, MySQL, Neo4j, Hive, HBase, Kafka, and JanusGraph.

Nebula Graph uses D3 for visualization. Three visualization toolkits are provided. Nebula Studio gives a graphical user interface to manipulate graph schema, import and navigate graph data, and run query statements. Nebula Explorer is designed as a no-code visual interaction with graph exploration. Nebula Dashboard can deploy and monitor the status of Nebula Graph on-premise or clouds.

\subsection{Observability} \label{sec:security}

Internal states of Nebula Graph are recorded and sent to a Prometheus service. Metrics like performance (insert/read/delete's QPS and latency), hardware usage (CPU, Memory, Disk, and Network) with different types (SUM, COUNT, AVG and p-quantiles), and time window (5 seconds to 1 hour) are collected. Google glog (run time log), folly xlog (audit log), and Linux logrotate are used to print and manage log files.

Like most DBMS, the execution plan is a frequently used and important way for the users to profile slow queries. The commands \textit{EXPLAIN} and \textit{PROFILE} are used to print optimized execution plans and show time cost for each execution plan node of a specific query statement. See an example in Fig.~\ref{fig:graphviz}. All the slow queries are sent to persist in the Meta Service by the Graph Service. And running slow queries can be killed by commands. All the logs and show queries can be sent to ELK frameworks for further analysis. 

\section{Benchmarking} \label{sec:bench}

We represent three benchmarking tests, two read tests and one write test.\footnote{Most graph databases are only publicly available for the single-server version and the distributed version is enterprise only. And LDBC gremlin's implementation is not available. Therefor we conduct benchmarking on Nebula Graph without any competitor's comparison.}
\subsection{Environment Setup}

All tests are running on three virtual machines with each of 8 vCPU (Intel Xeon Platinum 8260 CPU @ 2.30GHz), 32GB Mem, 1 TB SATA SSD, and 10 Gigabit Ethernet. The operating system is CentOS 7.9. 

On every server, we deploy all the Meta, Storage and Graph services (v3.1). All configurations are as default, except that KV-separation is setting $rocksdb\_enable\_kv\_separation=true$, and the partition number is set to 120.

The data set is generated by LDBC-SNB v0.3.3 with Scale Factor 300, i.e., about 300GB raw inputs. There are in total 817,316,105 vertices and 5,269,286,554 edges, with eight tags and 19 edge types.

The stress client is from a remote server. Every test contains some query sentences (group-query), and input query parameters (seeds). All the query sentences are written in either parameterized openCypher or nGQL sentences without any pre-compiling. All the requests are ad-hoc and responded synchronously. All client threads choose the seeds exclusively to run the same group-query. Therefore it is expected to be a shorter time when the threads increases. Every test runs three times to ensure consistent behavior.

\subsection{LDBC-SNB Short Reads}

There are seven \textit{Interactive Short (IS)} reads in LDBC-SNB Short Reads. That is IS-1 to IS-7 written in parameterized openCypher sentences.
\footnote{
The sentences can be found in \textrm{https://github.com/CPWstatic/ldbc\_snb\_interactive
\_implementations/tree/nebula0.3.3/nebula/queries/interactive-short-1.ngql} to \textrm{interactive-short-7.ngql}}
At the beginning of every group query, a seed is chosen, and the client sent IS-1 query to Nebula Graph and get the response. Then IS-2 query is sent and responded, etc. We sum all of the seven synchronous requests' latency. Then a new seed is chosen and a new sequence of IS-1 to IS-7 queries are sent, until all 10k seeds are done. It runs in 5, 10, 20, 30, 40, or 50 concurrent client threads to test the short depth read capability.

The QPS and latency are shown in Fig.~\Ref{fig:ldbc-is1} and hardware metrics in Fig.~\Ref{fig:ldbc-is2}. The \textit{accuracy} indicates that all requests are successful. It can be seen that the performance is CPU bounded at about 30 clients ($30\_vu$), and after that the latency goes up linearly. But before that, the QPS performance increases linearly with more stress. Besides, all servers' workloads are nearly equivalent.

\subsection{2-hop Breadth-First Search}

Breadth-first search is widely used in graph analytics for data collecting and recalling purposes. We test the specific clause \textrm{GO} by starting from a vertex 2-hops away and counting the neighboring numbers. The seed number is 10k and concurrent client threads are 5, 10, 20, 30, 40, or 50. An example query sentence is as below. 


\begin{minted}[frame=lines, framesep=2mm]{SQL} 
GO 2 STEPS FROM "p-10995116795962" OVER * 
YIELD DST(EDGE) | YIELD COUNT(*);
+----------+
| COUNT(*) |
+----------+
| 148742   |
+----------+
Got 1 rows (time spent 328233/329623 us)
\end{minted}

\begin{figure*}[htbp]
\centering
\subfigure{
\begin{minipage}[t]{0.3\linewidth}
\includegraphics[height=0.6\linewidth,width=1\linewidth]{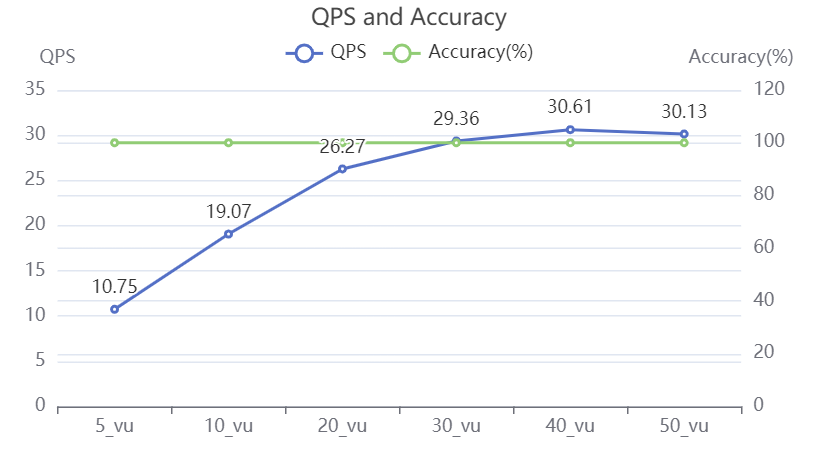}
\\
\includegraphics[height=0.6\linewidth,width=1\linewidth]{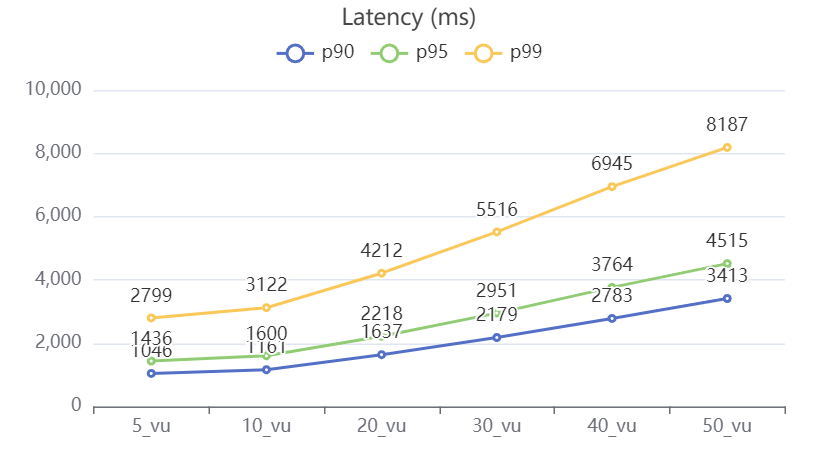}
\caption{QPS, Latency of LDBC-SNB Short Reads.}
\label{fig:ldbc-is1}
\end{minipage}
}
\subfigure{
\begin{minipage}[t]{0.3\linewidth}
\includegraphics[height=0.6\linewidth,width=1\linewidth]{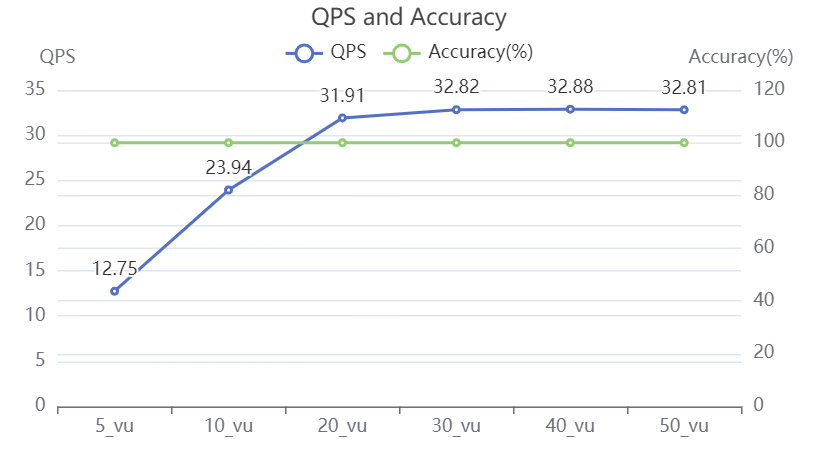}
\\
\includegraphics[height=0.6\linewidth,width=1\linewidth]{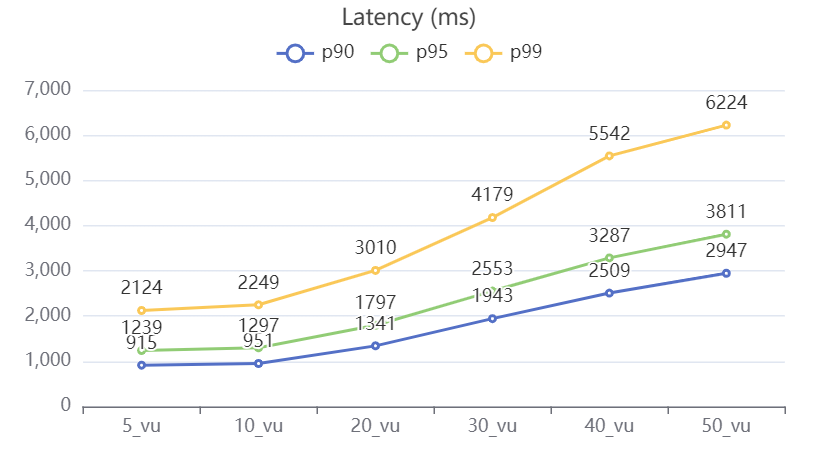}
\caption{QPS, Latency of 2-hop BFS.}
\label{fig:GO-is1}
\end{minipage}
}
\subfigure{
\begin{minipage}[t]{0.3\linewidth}
\includegraphics[height=0.6\linewidth,width=1\linewidth]{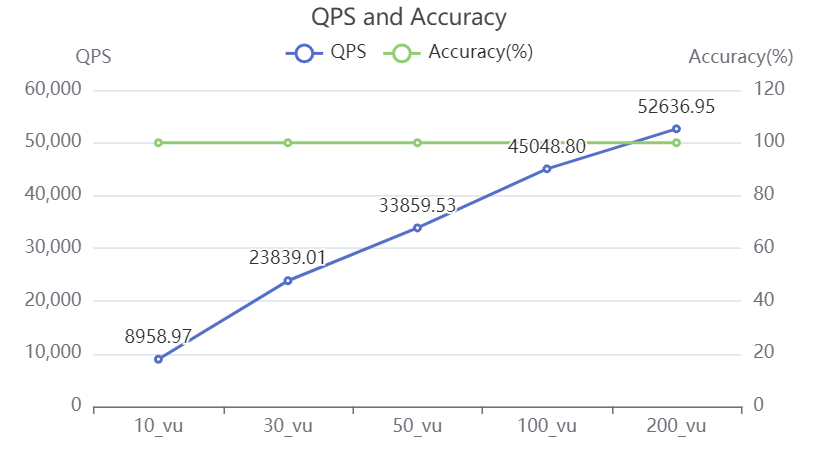}
\\
\includegraphics[height=0.6\linewidth,width=1\linewidth]{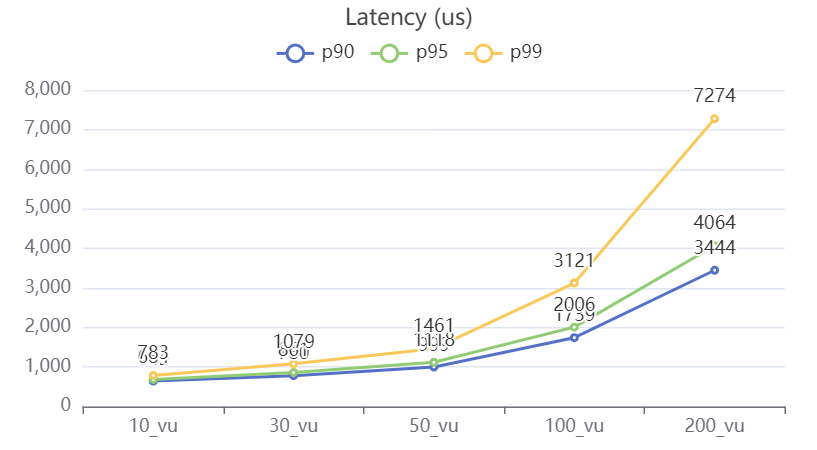}
\caption{QPS, Latency of Insert and Update.}
\label{fig:Insert-is1}
\end{minipage}
}

\subfigure{
\begin{minipage}[t]{0.3\linewidth}
\includegraphics[height=0.5\linewidth,width=1\linewidth]{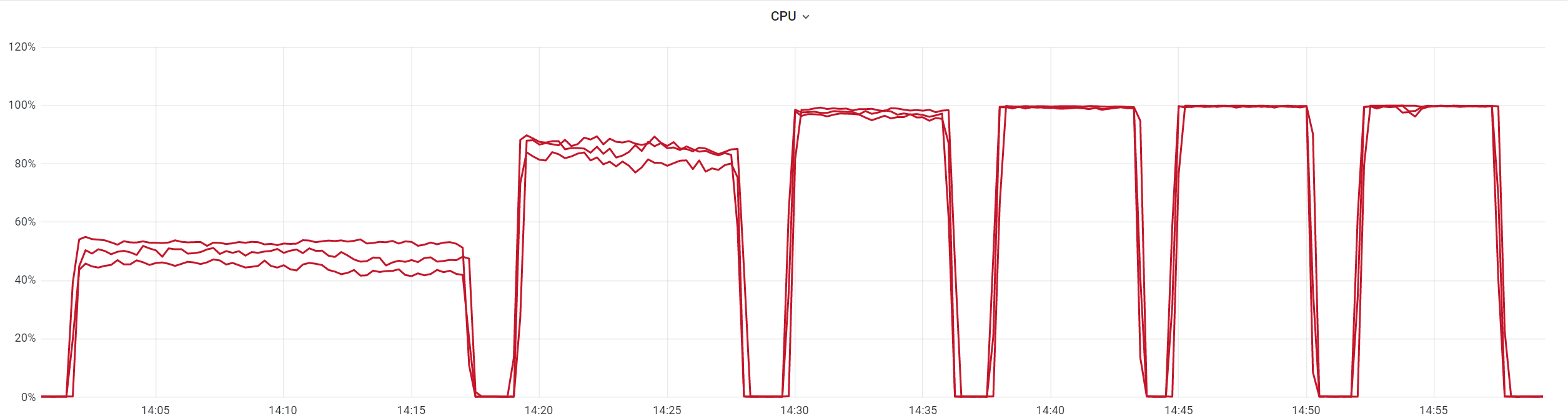}
\\
\includegraphics[height=0.5\linewidth,width=1\linewidth]{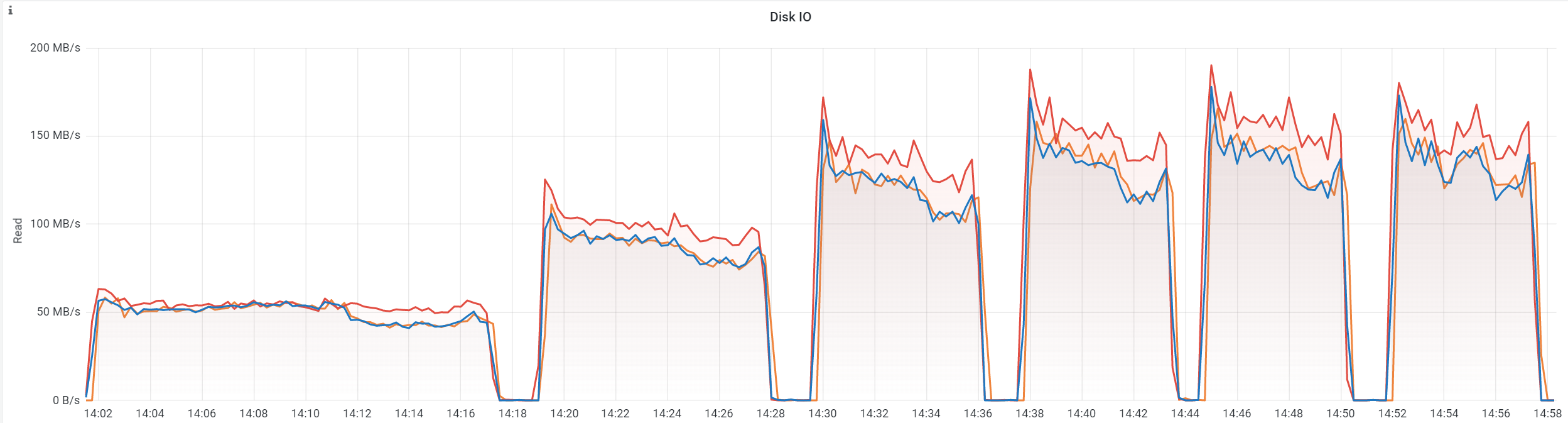}
\\
\includegraphics[height=0.5\linewidth,width=1\linewidth]{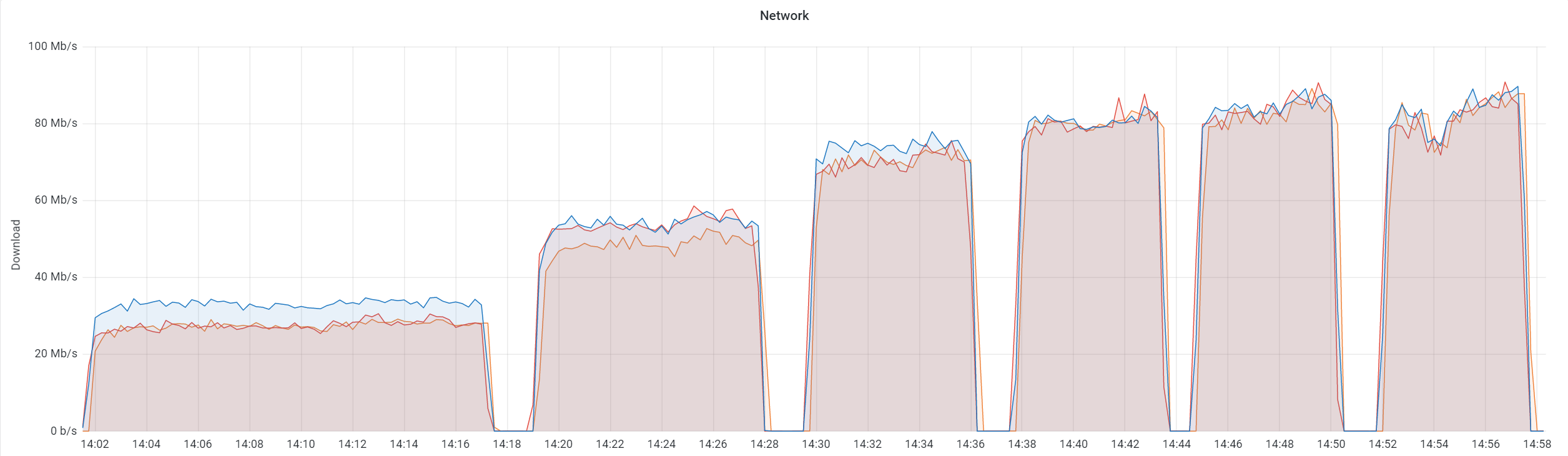}
\caption{Metric of CPU, Disk, and Network of LDBC Short Reads.}
\label{fig:ldbc-is2}
\end{minipage}
}
\subfigure{
\begin{minipage}[t]{0.3\linewidth}
\includegraphics[height=0.5\linewidth,width=1\linewidth]{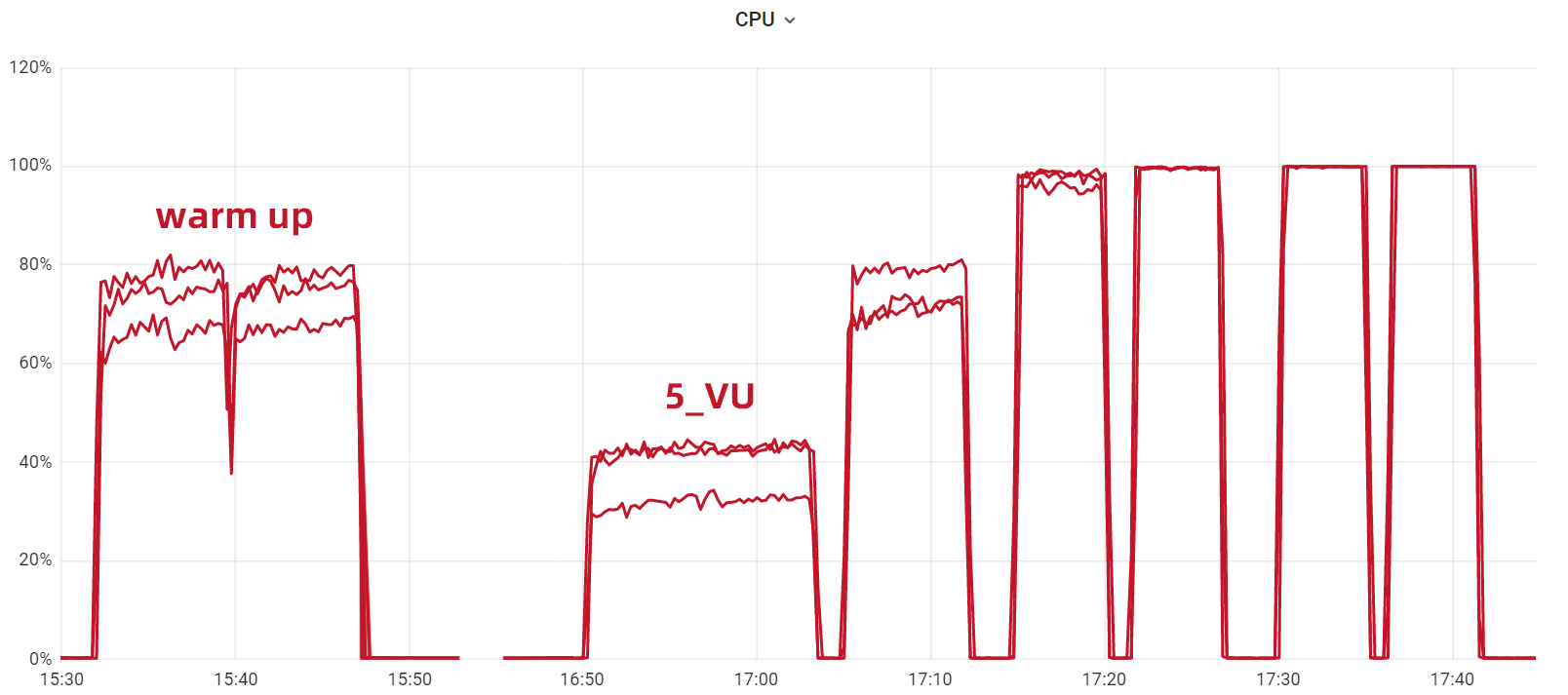}
\\
\includegraphics[height=0.5\linewidth,width=1\linewidth]{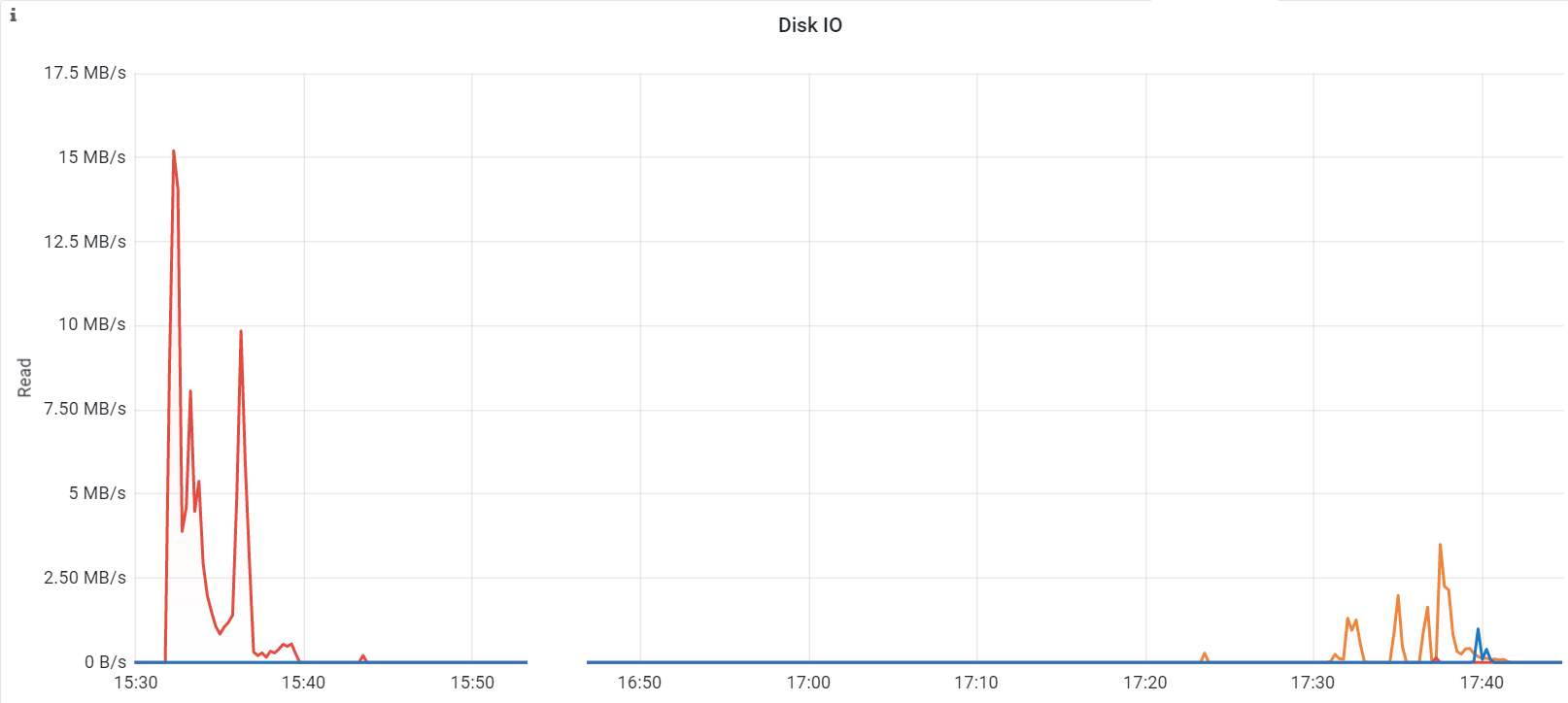}
\\
\includegraphics[height=0.5\linewidth,width=1\linewidth]{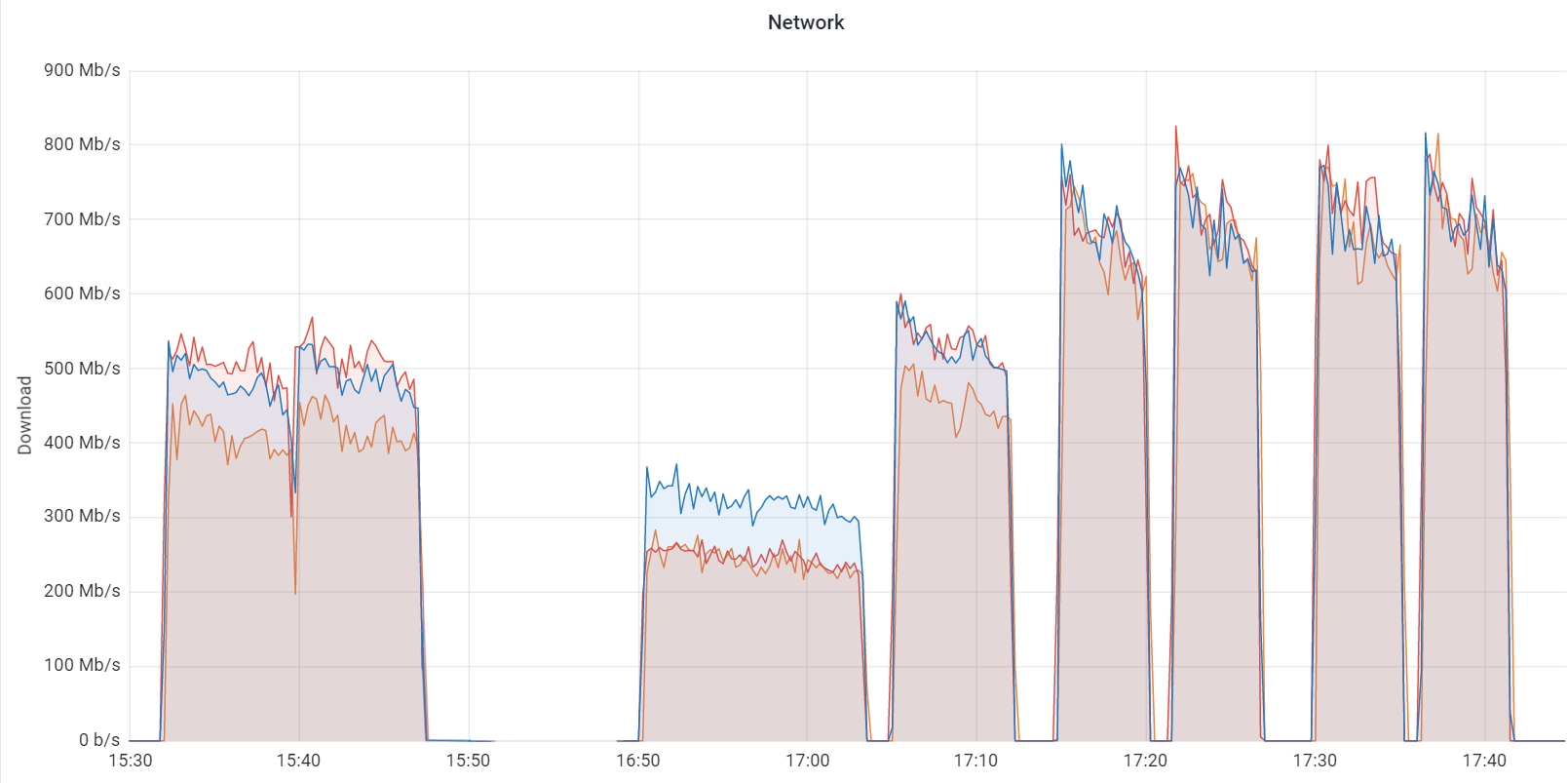}
\caption{Metric of CPU, Disk, and Network of 2-hop. The left side fluctuation is the warm-up round.}
\label{fig:GO-is2}
\end{minipage}
}
\subfigure{
\begin{minipage}[t]{0.3\linewidth}
\includegraphics[height=0.5\linewidth,width=1\linewidth]{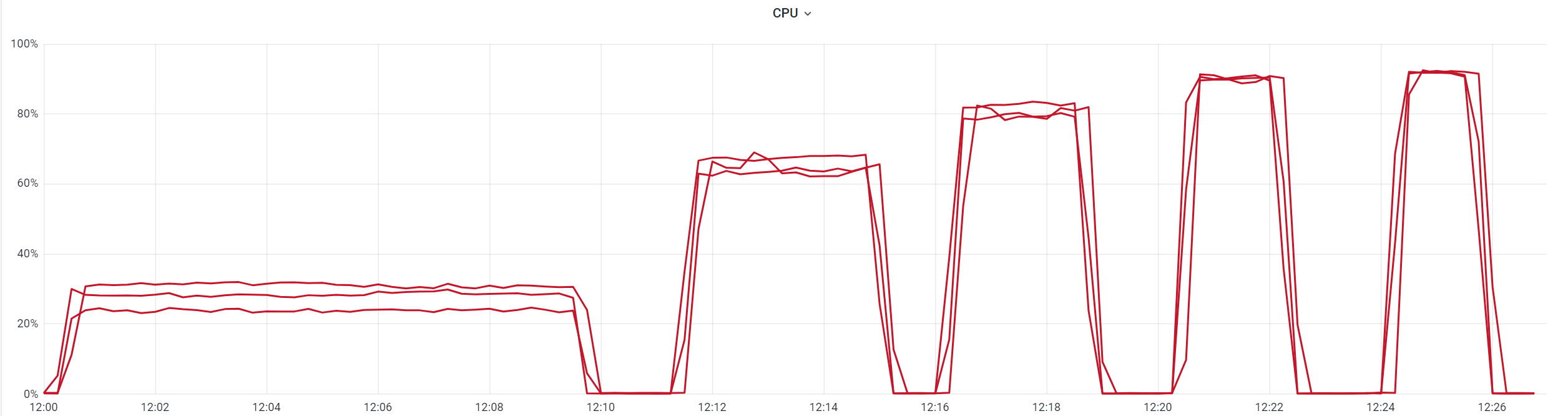}
\\
\includegraphics[height=0.5\linewidth,width=1\linewidth]{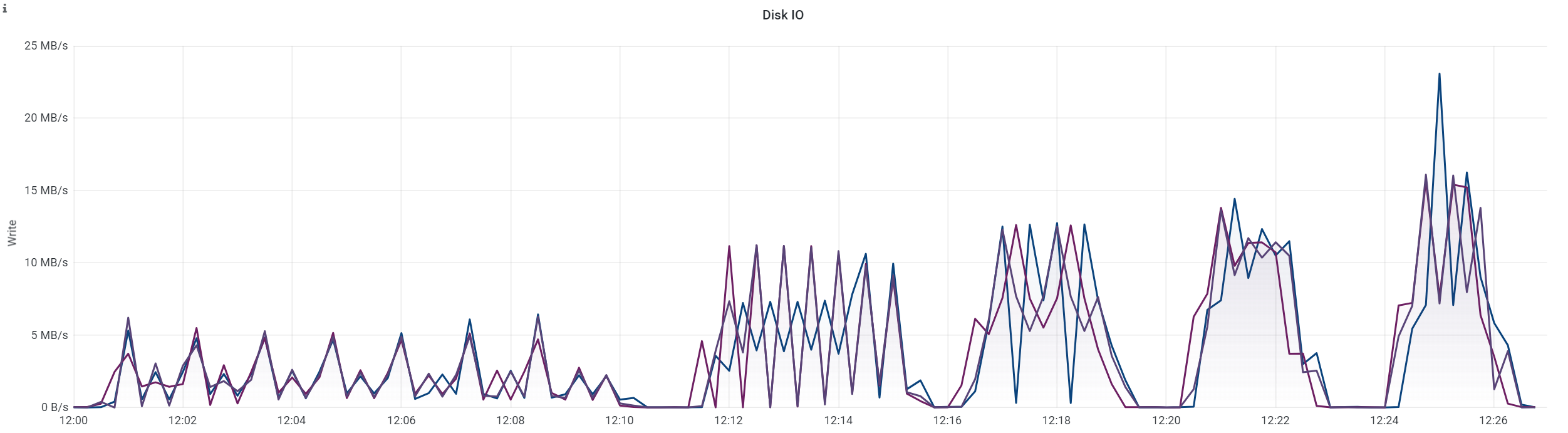}
\\
\includegraphics[height=0.5\linewidth,width=1\linewidth]{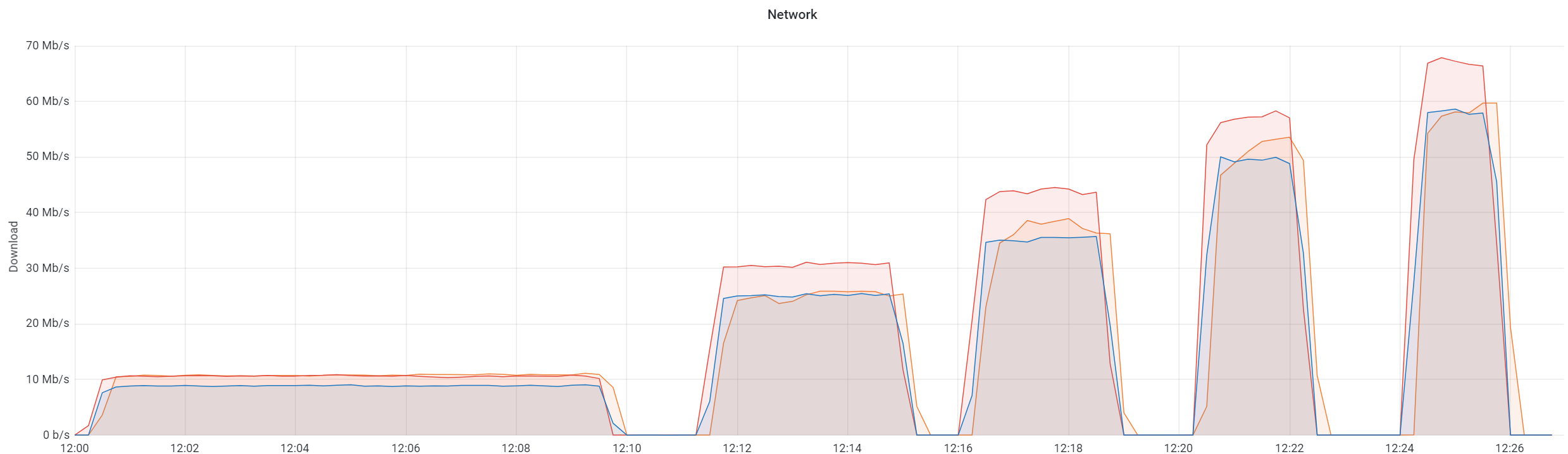}
\caption{Metric of QPS, Latency of Insert and Update.}
\label{fig:Insert-is2}
\end{minipage}
}

\end{figure*}

The test results are shown in Fig.\ref{fig:GO-is1} and \ref{fig:GO-is2}. Notice that the 2-hop search only requires the graph structure without any property. While the above Short Reads require both graph structure and property. Therefore KV-separation is much more useful in this case after warm-up, which is the disk reads pikes in Fig.~\ref{fig:GO-is2}. The QPS scales linearly as the stress increase until ~$30\_vu$ which then is CPU bounded.

\begin{figure*}[thp]
\includegraphics[width=\linewidth]{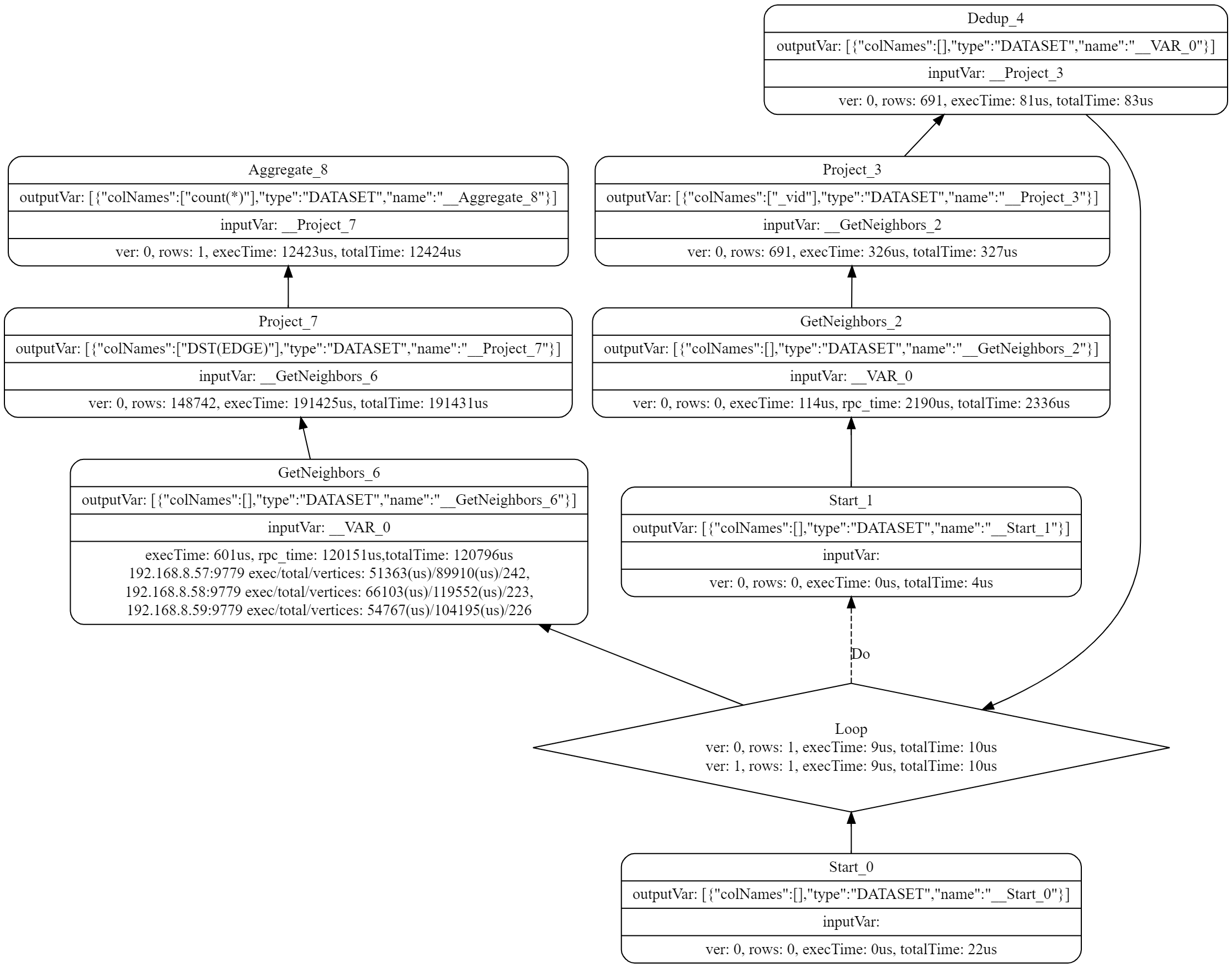}
\caption{Profiling of a sampling query of 2-hop.}
\label{fig:graphviz}
\end{figure*}

We further profile to see that the most time consuming steps (Fig.~\Ref{fig:graphviz}): \textit{GetNeighbors\_6} (120796us), which is to get vertices and edges from three Storage Services to one Graph Service through RPC, and \textit{Project\_7} (191431us), which is to project (choose) the DST(VID) column for every row since the Graph Service is row-based processing not columnar processing.


\subsection{LDBC-SNB Insert and Update}

\begin{figure*}[htp]
\includegraphics[width=\linewidth]{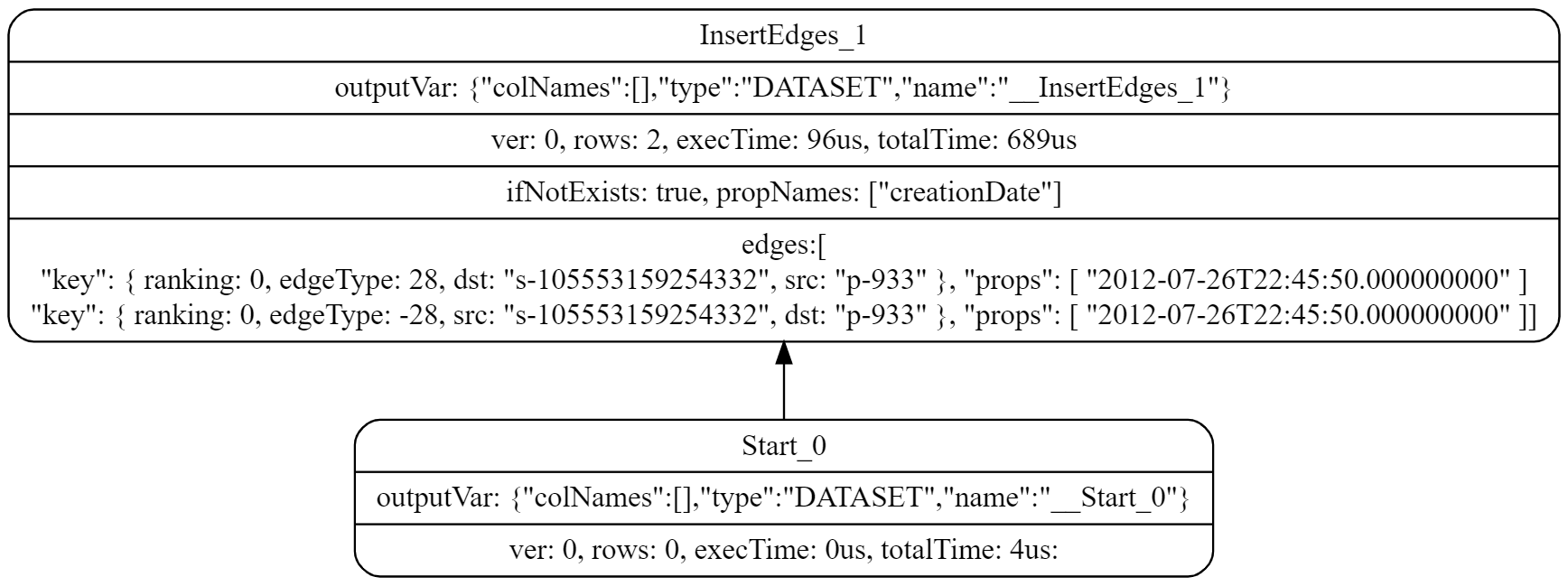}
\caption{Profiling of insert an edge.}
\label{fig:insertgraphviz}
\end{figure*}

The above two tests are used to stress the read performance. And this is to test the insert capability. The sentence is to insert a vertex or an edge with a given condition. E.g.,

\begin{minted}[frame=lines, framesep=2mm]{SQL} 
INSERT EDGE IF NOT EXISTS LIKES_POST(creationDate)
VALUES "p-933"->"s-105553159254332":
(datetime("2012-07-26T22:45:50"))
Execution succeeded (time spent 841/985 us)
\end{minted}

There are eight group-writes\footnote{
The implementation sentences can be found in \textrm{https://github.com/CPWstatic/ldbc\_snb\_interactive
\_implementations/tree/nebula0.3.3/nebula/queries/
interactive-update-1-add-person-companies.ngql} to \textrm{interactive-update-8.ngql}} in LDBC-SNB, IU-1 to IU-8. We choose five million seeds and run in 5, 10, 20, 30, 40, 50 concurrent clients. The results are shown in Fig.\ref{fig:Insert-is1} and \ref{fig:Insert-is2}. The execution plan of insertion is much simpler than reads, and the latency is also much shorter (Fig.~\ref{fig:insertgraphviz}). The fluctuation of disk IO is because of the RocksDB memtable flush. 

\section{Related Work}

In literature, there are numerous surveys on studying graph databases and related technologies, including graph languages, graph processing systems, and visualizations. Many evaluations and benchmarks are conducted as well. 

\textbf{Graph databases}, mainly designed for managing graph-like data, follow the principles of database systems, i.e. persistent storage, data consistency, execution plans, and query languages. The study of graph databases has a long history, at least since the 1970s\cite{DDIA}. We found many native graph databases are available: Neo4j\cite{guia2017graph,miller2013graph}, TigerGraph\cite{tiger1,tiger3}, JanusGraph\cite{JanusGraph}, RedisGraph\cite{cailliau2019redisgraph}, GeaBase\cite{Geabase}, GalaxyBase\cite{zhang2019platform}, and DGraph\cite{jain2005dgraph}. And some products are built directly upon other DBMSs via graph view or graph indexes, such as ArangoDB\cite{arangodb}, OrientDB\cite{tesoriero2013getting}, and Oracle Graph\cite{oraclegraphintro}.



\textbf{Graph Languages} Currently, there are three commonly used graph languages: Neo4j Cypher, Apache Gremlin, and W3C SPARQL; and some variants: Oracle PGQL\cite{pgql-lang}, TigerGraph GSQL\cite{tiger3}, and LDBC G-CORE\cite{angles2018g}. Researches on graph languages have bloomed in the last decade\cite{lindaaker2018overview,gql2017,gql2021,cypher,deutsch2021graph,marton2017formalising} with the emergence of graph databases. But these diversities of languages fragment the graph market from porting applications and skill sets from one platform to another. The ISO/IEC GQL standards have gained much progress, and after the release they are predicated to help all the vendors and clients adopt and converge on that language.  


\textbf{Graph Processing Frameworks}

In addition to graph databases, a number of graph processing frameworks have been proposed for large-scale graphs. These frameworks are characterized by in-memory batch processing and the use of distributed and parallel processing strategies. These frameworks perform iteratively and batch processes over the entire graph until the computation satisfies a fixed-point or stopping criterion. e.g., PageRank, clustering, machine learning, and data mining algorithms. Typical graph-specific platforms are Pregel\cite{malewicz2010pregel}, Apache Giraph\cite{sakr2016large}, GraphLab\cite{low2010graphlab}, PowerGraph\cite{gonzalez2012powergraph}, GraphX\cite{gonzalez2014graphx}, Graphscope\cite{fan2021graphscope}, and Plato\cite{tencentplato}. 

\textbf{Visualizations}

Graphs are welcome in data analysis and business intelligence because of the intuitive and heuristic thinking mode inspired by it. Most graph databases have their own visualization tools, like Neo4j Bloom\cite{bloom} and TigerGraph GraphStudio\cite{TigerGraph2022}. And some third-party libraries are available, e.g., D3\cite{vancak2017graph}, G6\cite{wang2021g6}, Cytoscape\cite{franz2016cytoscape}, Gephi\cite{cherven2013network}, and Zeppelin\cite{cheng2018building}.

\textbf{Graph Reachability and Structure Index}

The reachability problems on graph structure has been well studied\cite{Wei2014,jin2019shortest} and been applied in some graph databases. Except for kv-separation in Sec \Ref{sec:partition}  and secondary property indexes as in Sec \Ref{sec:index}, different types of indices are investigated to speed up structure based queries\cite{mhedhbi2021a+,10.1145/3282373.3282374,sumrall2016investigations,sakr2012overview,williams2007graph}.

\textbf{Comparison and Benchmarking}

Plenty of surveys and comparisons have been conducted among different graph databases\cite{besta2019demystifying,patil2018survey,besta2018survey,Angles2018,Sakr2021-dl,Sahu2020-pi,fernandes2018graph, meiling2017benchmarking}. 

In DBMS, TPC standards are widely accepted as performance benchmarks. But there are no globally accepted benchmarks in graph databases. Some research work has been done\cite{Macrobenchmarks,LDBC2020,LDBC-SNB,mhedhbi2021lsqb}, in which the most well-known benchmarking of a graph database is LDBC-SNB. Some third-party experiments are publicly available\cite{rusu2019depth,cailliau2019redisgraph,wangproc2020empriical}.

\section{Conclusions \& Future Work}

In this paper, we present Nebula Graph, an open-source graph database. Nebula Graph has already served in production in hundreds of leading companies, such as Walmart (WMT.NYSE), China Mobile (0941.HK), Tencent (0700.HK), Huawei, Meituan (3690.HK), JD (9618.HK), NetEase (NTES.NYSE), etc. The largest single cluster in production has a hundred servers, with more than 100 Terabytes.

In the future, three major works are planned. One is to be compatible with the upcoming ISO/IEC GQL standards, which requires a major redesign of the Graph Service, especially the optimizer. The second is to support running GQL in the ETL process of Nebula Core and Nebula Algorithms, which gives much flexibility by queries to get data instead of programming Scala codes in Spark. The third is to host Nebula Graph on Cloud, which requires a tremendous refactor of the Storage Service. 



\bibliography{sn-bibliography}


\begin{thebibliography}{71}
\ifx \bisbn   \undefined \def \bisbn  #1{ISBN #1}\fi
\ifx \binits  \undefined \def \binits#1{#1}\fi
\ifx \bauthor  \undefined \def \bauthor#1{#1}\fi
\ifx \batitle  \undefined \def \batitle#1{#1}\fi
\ifx \bjtitle  \undefined \def \bjtitle#1{#1}\fi
\ifx \bvolume  \undefined \def \bvolume#1{\textbf{#1}}\fi
\ifx \byear  \undefined \def \byear#1{#1}\fi
\ifx \bissue  \undefined \def \bissue#1{#1}\fi
\ifx \bfpage  \undefined \def \bfpage#1{#1}\fi
\ifx \blpage  \undefined \def \blpage #1{#1}\fi
\ifx \burl  \undefined \def \burl#1{\textsf{#1}}\fi
\ifx \doiurl  \undefined \def \doiurl#1{\url{https://doi.org/#1}}\fi
\ifx \betal  \undefined \def \betal{\textit{et al.}}\fi
\ifx \binstitute  \undefined \def \binstitute#1{#1}\fi
\ifx \binstitutionaled  \undefined \def \binstitutionaled#1{#1}\fi
\ifx \bctitle  \undefined \def \bctitle#1{#1}\fi
\ifx \beditor  \undefined \def \beditor#1{#1}\fi
\ifx \bpublisher  \undefined \def \bpublisher#1{#1}\fi
\ifx \bbtitle  \undefined \def \bbtitle#1{#1}\fi
\ifx \bedition  \undefined \def \bedition#1{#1}\fi
\ifx \bseriesno  \undefined \def \bseriesno#1{#1}\fi
\ifx \blocation  \undefined \def \blocation#1{#1}\fi
\ifx \bsertitle  \undefined \def \bsertitle#1{#1}\fi
\ifx \bsnm \undefined \def \bsnm#1{#1}\fi
\ifx \bsuffix \undefined \def \bsuffix#1{#1}\fi
\ifx \bparticle \undefined \def \bparticle#1{#1}\fi
\ifx \barticle \undefined \def \barticle#1{#1}\fi
\bibcommenthead
\ifx \bconfdate \undefined \def \bconfdate #1{#1}\fi
\ifx \botherref \undefined \def \botherref #1{#1}\fi
\ifx \url \undefined \def \url#1{\textsf{#1}}\fi
\ifx \bchapter \undefined \def \bchapter#1{#1}\fi
\ifx \bbook \undefined \def \bbook#1{#1}\fi
\ifx \bcomment \undefined \def \bcomment#1{#1}\fi
\ifx \oauthor \undefined \def \oauthor#1{#1}\fi
\ifx \citeauthoryear \undefined \def \citeauthoryear#1{#1}\fi
\ifx \endbibitem  \undefined \def \endbibitem {}\fi
\ifx \bconflocation  \undefined \def \bconflocation#1{#1}\fi
\ifx \arxivurl  \undefined \def \arxivurl#1{\textsf{#1}}\fi
\csname PreBibitemsHook\endcsname

\bibitem{dbengineranking}
\begin{botherref}
\oauthor{\bsnm{{Solid IT gmbh}}}:
Complete trend, starting with January 2013
(2022).
\url{https://db-engines.com/en/ranking\_categories}
\end{botherref}
\endbibitem

\bibitem{ongaro2015Raft}
\begin{botherref}
\oauthor{\bsnm{Ongaro}, \binits{D.}},
\oauthor{\bsnm{Ousterhout}, \binits{J.}}:
The raft consensus algorithm
(2015)
\end{botherref}
\endbibitem

\bibitem{k8s}
\begin{bbook}
\bauthor{\bsnm{Sayfan}, \binits{G.}}:
\bbtitle{Mastering Kubernetes}.
\bpublisher{Packt Publishing Ltd},
\blocation{Birmingham}
(\byear{2017})
\end{bbook}
\endbibitem

\bibitem{rocksdb}
\begin{bchapter}
\bauthor{\bsnm{Cao}, \binits{Z.}},
\bauthor{\bsnm{Dong}, \binits{S.}},
\bauthor{\bsnm{Vemuri}, \binits{S.}},
\bauthor{\bsnm{Du}, \binits{D.H.}}:
\bctitle{Characterizing, modeling, and benchmarking rocksdb key-value workloads
  at facebook}.
In: \bbtitle{18th USENIX Conference on File and Storage Technologies (FAST
  20)},
pp. \bfpage{209}--\blpage{223}
(\byear{2020})
\end{bchapter}
\endbibitem

\bibitem{cypher}
\begin{bchapter}
\bauthor{\bsnm{Francis}, \binits{N.}},
\bauthor{\bsnm{Green}, \binits{A.}},
\bauthor{\bsnm{Guagliardo}, \binits{P.}},
\bauthor{\bsnm{Libkin}, \binits{L.}},
\bauthor{\bsnm{Lindaaker}, \binits{T.}},
\bauthor{\bsnm{Marsault}, \binits{V.}},
\bauthor{\bsnm{Plantikow}, \binits{S.}},
\bauthor{\bsnm{Rydberg}, \binits{M.}},
\bauthor{\bsnm{Selmer}, \binits{P.}},
\bauthor{\bsnm{Taylor}, \binits{A.}}:
\bctitle{Cypher: An evolving query language for property graphs}.
In: \bbtitle{Proceedings of the 2018 International Conference on Management of
  Data}.
\bsertitle{SIGMOD '18},
pp. \bfpage{1433}--\blpage{1445}.
\bpublisher{Association for Computing Machinery},
\blocation{New York, NY, USA}
(\byear{2018}).
\burl{https://doi.org/10.1145/3183713.3190657}
\end{bchapter}
\endbibitem

\bibitem{neo4jsite}
\begin{botherref}
\oauthor{\bsnm{{Neo4j, Inc.}}}:
NEO4J GRAPH DATA PLATFORM.
Accessed: March 28, 2022
(2022).
\url{https://neo4j.com/}
\end{botherref}
\endbibitem

\bibitem{gql2017}
\begin{botherref}
\oauthor{\bsnm{Angles}, \binits{R.}},
\oauthor{\bsnm{Arenas}, \binits{M.}},
\oauthor{\bsnm{Barcel\'{o}}, \binits{P.}},
\oauthor{\bsnm{Hogan}, \binits{A.}},
\oauthor{\bsnm{Reutter}, \binits{J.}},
\oauthor{\bsnm{Vrgo\v{c}}, \binits{D.}}:
Foundations of modern query languages for graph databases.
ACM Comput. Surv.
\textbf{50}(5)
(2017).
\doiurl{10.1145/3104031}
\end{botherref}
\endbibitem

\bibitem{lu2017wisckey}
\begin{barticle}
\bauthor{\bsnm{Lu}, \binits{L.}},
\bauthor{\bsnm{Pillai}, \binits{T.S.}},
\bauthor{\bsnm{Gopalakrishnan}, \binits{H.}},
\bauthor{\bsnm{Arpaci-Dusseau}, \binits{A.C.}},
\bauthor{\bsnm{Arpaci-Dusseau}, \binits{R.H.}}:
\batitle{Wisckey: Separating keys from values in ssd-conscious storage}.
\bjtitle{ACM Transactions on Storage (TOS)}
\bvolume{13}(\bissue{1}),
\bfpage{1}--\blpage{28}
(\byear{2017})
\end{barticle}
\endbibitem

\bibitem{gormley2015elasticsearch}
\begin{bbook}
\bauthor{\bsnm{Gormley}, \binits{C.}},
\bauthor{\bsnm{Tong}, \binits{Z.}}:
\bbtitle{Elasticsearch: the Definitive Guide: a Distributed Real-time Search
  and Analytics Engine}.
\bpublisher{" O'Reilly Media, Inc."},
\blocation{New York}
(\byear{2015})
\end{bbook}
\endbibitem

\bibitem{googles2}
\begin{botherref}
\oauthor{\bsnm{Google}}:
S2 Geometry
(2022).
\url{https://s2geometry.io/}
\end{botherref}
\endbibitem

\bibitem{LDBC-SNB}
\begin{bchapter}
\bauthor{\bsnm{Erling}, \binits{O.}},
\bauthor{\bsnm{Averbuch}, \binits{A.}},
\bauthor{\bsnm{Larriba-Pey}, \binits{J.}},
\bauthor{\bsnm{Chafi}, \binits{H.}},
\bauthor{\bsnm{Gubichev}, \binits{A.}},
\bauthor{\bsnm{Prat}, \binits{A.}},
\bauthor{\bsnm{Pham}, \binits{M.-D.}},
\bauthor{\bsnm{Boncz}, \binits{P.}}:
\bctitle{The ldbc social network benchmark: Interactive workload}.
In: \bbtitle{Proceedings of the 2015 ACM SIGMOD International Conference on
  Management of Data}.
\bsertitle{SIGMOD '15},
pp. \bfpage{619}--\blpage{630}.
\bpublisher{Association for Computing Machinery},
\blocation{New York, NY, USA}
(\byear{2015}).
\burl{https://doi.org/10.1145/2723372.2742786}
\end{bchapter}
\endbibitem

\bibitem{graefe1995cascades}
\begin{barticle}
\bauthor{\bsnm{Graefe}, \binits{G.}}:
\batitle{The cascades framework for query optimization}.
\bjtitle{IEEE Data Eng. Bull.}
\bvolume{18}(\bissue{3}),
\bfpage{19}--\blpage{29}
(\byear{1995})
\end{barticle}
\endbibitem

\bibitem{escriva2015warp}
\begin{botherref}
\oauthor{\bsnm{Escriva}, \binits{R.}},
\oauthor{\bsnm{Wong}, \binits{B.}},
\oauthor{\bsnm{Sirer}, \binits{E.G.}}:
Warp: Lightweight multi-key transactions for key-value stores.
arXiv preprint arXiv:1509.07815
(2015)
\end{botherref}
\endbibitem

\bibitem{NebulaManual}
\begin{botherref}
\oauthor{\bsnm{Wu}, \binits{M.}},
\oauthor{\bsnm{Zhou}, \binits{Y.}},
\oauthor{\bsnm{Liang}, \binits{C.}},
\oauthor{\bsnm{Yang}, \binits{F.}},
\oauthor{\bsnm{Huang}, \binits{A.}}:
Nebula Graph Database Manual
(2022).
\url{https://docs.nebula-graph.io/3.0.2/pdf/NebulaGraph-EN.pdf}
\end{botherref}
\endbibitem

\bibitem{rodriguez2015gremlin}
\begin{bchapter}
\bauthor{\bsnm{Rodriguez}, \binits{M.A.}}:
\bctitle{The gremlin graph traversal machine and language (invited talk)}.
In: \bbtitle{Proceedings of the 15th Symposium on Database Programming
  Languages},
pp. \bfpage{1}--\blpage{10}
(\byear{2015})
\end{bchapter}
\endbibitem

\bibitem{world2013sparql}
\begin{botherref}
\oauthor{\bsnm{Consortium}, \binits{W.W.W.}}, et al.:
Sparql 1.1 overview
(2013)
\end{botherref}
\endbibitem

\bibitem{openCypherTCK}
\begin{botherref}
\oauthor{\bsnm{{Neo4j Inc.}}}:
Technology Compatibility Kit (TCK), openCypher Resources
(2021).
\url{https://s3.amazonaws.com/artifacts.opencypher.org/M18/tck-M18.zip}
\end{botherref}
\endbibitem

\bibitem{gonzalez2012powergraph}
\begin{bchapter}
\bauthor{\bsnm{Gonzalez}, \binits{J.E.}},
\bauthor{\bsnm{Low}, \binits{Y.}},
\bauthor{\bsnm{Gu}, \binits{H.}},
\bauthor{\bsnm{Bickson}, \binits{D.}},
\bauthor{\bsnm{Guestrin}, \binits{C.}}:
\bctitle{$\{$PowerGraph$\}$: Distributed $\{$Graph-Parallel$\}$ computation on
  natural graphs}.
In: \bbtitle{10th USENIX Symposium on Operating Systems Design and
  Implementation (OSDI 12)},
pp. \bfpage{17}--\blpage{30}
(\byear{2012})
\end{bchapter}
\endbibitem

\bibitem{gonzalez2014graphx}
\begin{bchapter}
\bauthor{\bsnm{Gonzalez}, \binits{J.E.}},
\bauthor{\bsnm{Xin}, \binits{R.S.}},
\bauthor{\bsnm{Dave}, \binits{A.}},
\bauthor{\bsnm{Crankshaw}, \binits{D.}},
\bauthor{\bsnm{Franklin}, \binits{M.J.}},
\bauthor{\bsnm{Stoica}, \binits{I.}}:
\bctitle{$\{$GraphX$\}$: Graph processing in a distributed dataflow framework}.
In: \bbtitle{11th USENIX Symposium on Operating Systems Design and
  Implementation (OSDI 14)},
pp. \bfpage{599}--\blpage{613}
(\byear{2014})
\end{bchapter}
\endbibitem

\bibitem{tencentplato}
\begin{botherref}
\oauthor{\bsnm{Tencent}}:
Tencent Graph Computing (TGraph) Officially Open Sourced High-Performance Graph
  Computing Framework: Plato.
2022
(2022).
\url{https://github.com/Tencent/plato/blob/master/doc/introduction\_en.md}
\end{botherref}
\endbibitem

\bibitem{watson2014under}
\begin{botherref}
\oauthor{\bsnm{Watson}, \binits{D.}}:
Under the hood: Building and open-sourcing fbthrift
(2014)
\end{botherref}
\endbibitem

\bibitem{DDIA}
\begin{bbook}
\bauthor{\bsnm{Kleppmann}, \binits{M.}}:
\bbtitle{Designing Data-intensive Applications: The Big Ideas Behind Reliable,
  Scalable, and Maintainable Systems}.
\bpublisher{O'Reilly Media, Inc.},
\blocation{Sebastopol, CA}
(\byear{2017})
\end{bbook}
\endbibitem

\bibitem{guia2017graph}
\begin{bchapter}
\bauthor{\bsnm{Guia}, \binits{J.}},
\bauthor{\bsnm{Soares}, \binits{V.G.}},
\bauthor{\bsnm{Bernardino}, \binits{J.}}:
\bctitle{Graph databases: Neo4j analysis.}
In: \bbtitle{ICEIS (1)},
pp. \bfpage{351}--\blpage{356}
(\byear{2017})
\end{bchapter}
\endbibitem

\bibitem{miller2013graph}
\begin{bchapter}
\bauthor{\bsnm{Miller}, \binits{J.J.}}:
\bctitle{Graph database applications and concepts with neo4j}.
In: \bbtitle{Proceedings of the Southern Association for Information Systems
  Conference, Atlanta, GA, USA},
vol. \bseriesno{2324}
(\byear{2013})
\end{bchapter}
\endbibitem

\bibitem{tiger1}
\begin{botherref}
\oauthor{\bsnm{Deutsch}, \binits{A.}},
\oauthor{\bsnm{Xu}, \binits{Y.}},
\oauthor{\bsnm{Wu}, \binits{M.}},
\oauthor{\bsnm{Lee}, \binits{V.}}:
TigerGraph: A Native MPP Graph Database.
arXiv
(2019).
\doiurl{10.48550/ARXIV.1901.08248}
\end{botherref}
\endbibitem

\bibitem{tiger3}
\begin{bchapter}
\bauthor{\bsnm{Deutsch}, \binits{A.}},
\bauthor{\bsnm{Xu}, \binits{Y.}},
\bauthor{\bsnm{Wu}, \binits{M.}},
\bauthor{\bsnm{Lee}, \binits{V.E.}}:
\bctitle{Aggregation support for modern graph analytics in tigergraph}.
In: \bbtitle{Proceedings of the 2020 ACM SIGMOD International Conference on
  Management of Data},
pp. \bfpage{377}--\blpage{392}
(\byear{2020})
\end{bchapter}
\endbibitem

\bibitem{JanusGraph}
\begin{botherref}
\oauthor{\bsnm{{JanusGraph Authors}}}:
JanusGraph: Distributed, open source, massively scalable graph database.
Accessed: March 28, 2022
(2022).
\url{https://janusgraph.org/}
\end{botherref}
\endbibitem

\bibitem{cailliau2019redisgraph}
\begin{bchapter}
\bauthor{\bsnm{Cailliau}, \binits{P.}},
\bauthor{\bsnm{Davis}, \binits{T.}},
\bauthor{\bsnm{Gadepally}, \binits{V.}},
\bauthor{\bsnm{Kepner}, \binits{J.}},
\bauthor{\bsnm{Lipman}, \binits{R.}},
\bauthor{\bsnm{Lovitz}, \binits{J.}},
\bauthor{\bsnm{Ouaknine}, \binits{K.}}:
\bctitle{Redisgraph graphblas enabled graph database}.
In: \bbtitle{2019 IEEE International Parallel and Distributed Processing
  Symposium Workshops (IPDPSW)},
pp. \bfpage{285}--\blpage{286}
(\byear{2019}).
\bcomment{IEEE}
\end{bchapter}
\endbibitem

\bibitem{Geabase}
\begin{bchapter}
\bauthor{\bsnm{Fu}, \binits{Z.}},
\bauthor{\bsnm{Wu}, \binits{Z.}},
\bauthor{\bsnm{Li}, \binits{H.}},
\bauthor{\bsnm{Li}, \binits{Y.}},
\bauthor{\bsnm{Wu}, \binits{M.}},
\bauthor{\bsnm{Chen}, \binits{X.}},
\bauthor{\bsnm{Ye}, \binits{X.}},
\bauthor{\bsnm{Yu}, \binits{B.}},
\bauthor{\bsnm{Hu}, \binits{X.}}:
\bctitle{Geabase: A high-performance distributed graph database for
  industry-scale applications}.
In: \bbtitle{2017 Fifth International Conference on Advanced Cloud and Big Data
  (CBD)},
pp. \bfpage{170}--\blpage{175}
(\byear{2017}).
\doiurl{10.1109/CBD.2017.37}
\end{bchapter}
\endbibitem

\bibitem{zhang2019platform}
\begin{botherref}
\oauthor{\bsnm{Zhang}, \binits{C.}},
\oauthor{\bsnm{Wu}, \binits{J.}}:
Platform, system, process for distributed graph databases and computing.
Google Patents.
US Patent 10,409,782
(2019)
\end{botherref}
\endbibitem

\bibitem{jain2005dgraph}
\begin{botherref}
\oauthor{\bsnm{Jain}, \binits{M.}}:
Dgraph: Synchronously replicated, transactional and distributed graph database
(2021)
\end{botherref}
\endbibitem

\bibitem{arangodb}
\begin{botherref}
\oauthor{\bsnm{{ArangoDB Inc}}}:
Arangodb
(2022).
\url{https://www.arangodb.com/}
\end{botherref}
\endbibitem

\bibitem{tesoriero2013getting}
\begin{bbook}
\bauthor{\bsnm{Tesoriero}, \binits{C.}}:
\bbtitle{Getting Started with OrientDB}.
\bpublisher{Packt Publishing Ltd},
\blocation{Birmingham}
(\byear{2013})
\end{bbook}
\endbibitem

\bibitem{oraclegraphintro}
\begin{botherref}
\oauthor{\bsnm{Corporation}, \binits{O.}}:
Graph Database and Graph Analytics
(2022).
\url{https://www.oracle.com/database/graph/}
\end{botherref}
\endbibitem

\bibitem{pgql-lang}
\begin{botherref}
\oauthor{\bsnm{Oracle}}:
Property Graph Query Language (PGQL).
Accessed: March 28, 2022
(2022).
\url{https://pgql-lang.org/}
\end{botherref}
\endbibitem

\bibitem{angles2018g}
\begin{bchapter}
\bauthor{\bsnm{Angles}, \binits{R.}},
\bauthor{\bsnm{Arenas}, \binits{M.}},
\bauthor{\bsnm{Barcel{\'o}}, \binits{P.}},
\bauthor{\bsnm{Boncz}, \binits{P.}},
\bauthor{\bsnm{Fletcher}, \binits{G.}},
\bauthor{\bsnm{Gutierrez}, \binits{C.}},
\bauthor{\bsnm{Lindaaker}, \binits{T.}},
\bauthor{\bsnm{Paradies}, \binits{M.}},
\bauthor{\bsnm{Plantikow}, \binits{S.}},
\bauthor{\bsnm{Sequeda}, \binits{J.}}, \betal:
\bctitle{G-core: A core for future graph query languages}.
In: \bbtitle{Proceedings of the 2018 International Conference on Management of
  Data},
pp. \bfpage{1421}--\blpage{1432}
(\byear{2018})
\end{bchapter}
\endbibitem

\bibitem{lindaaker2018overview}
\begin{botherref}
\oauthor{\bsnm{Lindaaker}, \binits{T.}}:
An overview of the recent history of graph query languages
(2018)
\end{botherref}
\endbibitem

\bibitem{gql2021}
\begin{bchapter}
\bauthor{\bsnm{Angles}, \binits{R.}},
\bauthor{\bsnm{Bonifati}, \binits{A.}},
\bauthor{\bsnm{Dumbrava}, \binits{S.}},
\bauthor{\bsnm{Fletcher}, \binits{G.}},
\bauthor{\bsnm{Hare}, \binits{K.W.}},
\bauthor{\bsnm{Hidders}, \binits{J.}},
\bauthor{\bsnm{Lee}, \binits{V.E.}},
\bauthor{\bsnm{Li}, \binits{B.}},
\bauthor{\bsnm{Libkin}, \binits{L.}},
\bauthor{\bsnm{Martens}, \binits{W.}},
\bauthor{\bsnm{Murlak}, \binits{F.}},
\bauthor{\bsnm{Perryman}, \binits{J.}},
\bauthor{\bsnm{Savkovi\'{c}}, \binits{O.}},
\bauthor{\bsnm{Schmidt}, \binits{M.}},
\bauthor{\bsnm{Sequeda}, \binits{J.}},
\bauthor{\bsnm{Staworko}, \binits{S.}},
\bauthor{\bsnm{Tomaszuk}, \binits{D.}}:
\bctitle{Pg-keys: Keys for property graphs}.
In: \bbtitle{Proceedings of the 2021 International Conference on Management of
  Data}.
\bsertitle{SIGMOD/PODS '21},
pp. \bfpage{2423}--\blpage{2436}.
\bpublisher{Association for Computing Machinery},
\blocation{New York, NY, USA}
(\byear{2021}).
\doiurl{10.1145/3448016.3457561}
\end{bchapter}
\endbibitem

\bibitem{deutsch2021graph}
\begin{botherref}
\oauthor{\bsnm{Deutsch}, \binits{A.}},
\oauthor{\bsnm{Francis}, \binits{N.}},
\oauthor{\bsnm{Green}, \binits{A.}},
\oauthor{\bsnm{Hare}, \binits{K.}},
\oauthor{\bsnm{Li}, \binits{B.}},
\oauthor{\bsnm{Libkin}, \binits{L.}},
\oauthor{\bsnm{Lindaaker}, \binits{T.}},
\oauthor{\bsnm{Marsault}, \binits{V.}},
\oauthor{\bsnm{Martens}, \binits{W.}},
\oauthor{\bsnm{Michels}, \binits{J.}}, et al.:
Graph pattern matching in gql and sql/pgq.
arXiv preprint arXiv:2112.06217
(2021)
\end{botherref}
\endbibitem

\bibitem{marton2017formalising}
\begin{bchapter}
\bauthor{\bsnm{Marton}, \binits{J.}},
\bauthor{\bsnm{Sz{\'a}rnyas}, \binits{G.}},
\bauthor{\bsnm{Varr{\'o}}, \binits{D.}}:
\bctitle{Formalising opencypher graph queries in relational algebra}.
In: \bbtitle{European Conference on Advances in Databases and Information
  Systems},
pp. \bfpage{182}--\blpage{196}
(\byear{2017}).
\bcomment{Springer}
\end{bchapter}
\endbibitem

\bibitem{malewicz2010pregel}
\begin{bchapter}
\bauthor{\bsnm{Malewicz}, \binits{G.}},
\bauthor{\bsnm{Austern}, \binits{M.H.}},
\bauthor{\bsnm{Bik}, \binits{A.J.}},
\bauthor{\bsnm{Dehnert}, \binits{J.C.}},
\bauthor{\bsnm{Horn}, \binits{I.}},
\bauthor{\bsnm{Leiser}, \binits{N.}},
\bauthor{\bsnm{Czajkowski}, \binits{G.}}:
\bctitle{Pregel: a system for large-scale graph processing}.
In: \bbtitle{Proceedings of the 2010 ACM SIGMOD International Conference on
  Management of Data},
pp. \bfpage{135}--\blpage{146}
(\byear{2010})
\end{bchapter}
\endbibitem

\bibitem{sakr2016large}
\begin{bbook}
\bauthor{\bsnm{Sakr}, \binits{S.}},
\bauthor{\bsnm{Orakzai}, \binits{F.M.}},
\bauthor{\bsnm{Abdelaziz}, \binits{I.}},
\bauthor{\bsnm{Khayyat}, \binits{Z.}}:
\bbtitle{Large-scale Graph Processing Using Apache Giraph}.
\bpublisher{Springer},
\blocation{New York}
(\byear{2016})
\end{bbook}
\endbibitem

\bibitem{low2010graphlab}
\begin{bchapter}
\bauthor{\bsnm{Low}, \binits{Y.}},
\bauthor{\bsnm{Gonzalez}, \binits{J.}},
\bauthor{\bsnm{Kyrola}, \binits{A.}},
\bauthor{\bsnm{Bickson}, \binits{D.}},
\bauthor{\bsnm{Guestrin}, \binits{C.}},
\bauthor{\bsnm{Hellerstein}, \binits{J.M.}}:
\bctitle{Graphlab: A new parallel framework for machine learning}.
In: \bbtitle{Conference on Uncertainty in Artificial Intelligence (UAI)},
vol. \bseriesno{20}
(\byear{2010})
\end{bchapter}
\endbibitem

\bibitem{fan2021graphscope}
\begin{barticle}
\bauthor{\bsnm{Fan}, \binits{W.}},
\bauthor{\bsnm{He}, \binits{T.}},
\bauthor{\bsnm{Lai}, \binits{L.}},
\bauthor{\bsnm{Li}, \binits{X.}},
\bauthor{\bsnm{Li}, \binits{Y.}},
\bauthor{\bsnm{Li}, \binits{Z.}},
\bauthor{\bsnm{Qian}, \binits{Z.}},
\bauthor{\bsnm{Tian}, \binits{C.}},
\bauthor{\bsnm{Wang}, \binits{L.}},
\bauthor{\bsnm{Xu}, \binits{J.}}, \betal:
\batitle{Graphscope: a unified engine for big graph processing}.
\bjtitle{Proceedings of the VLDB Endowment}
\bvolume{14}(\bissue{12}),
\bfpage{2879}--\blpage{2892}
(\byear{2021})
\end{barticle}
\endbibitem

\bibitem{bloom}
\begin{botherref}
\oauthor{\bsnm{{Neo4j, Inc.}}}:
Neo4j Bloom
(2022).
\url{https://neo4j.com/product/bloom/}
\end{botherref}
\endbibitem

\bibitem{TigerGraph2022}
\begin{botherref}
\oauthor{\bsnm{TigerGraph}}:
GraphStudio User Interface
(2022).
\url{https://www.tigergraph.com/graphstudio/}
\end{botherref}
\endbibitem

\bibitem{vancak2017graph}
\begin{botherref}
\oauthor{\bsnm{Vanc{\'a}k}, \binits{V.}}:
Graph data visualizations with d3. js libraries
(2017)
\end{botherref}
\endbibitem

\bibitem{wang2021g6}
\begin{barticle}
\bauthor{\bsnm{Wang}, \binits{Y.}},
\bauthor{\bsnm{Bai}, \binits{Z.}},
\bauthor{\bsnm{Lin}, \binits{Z.}},
\bauthor{\bsnm{Dong}, \binits{X.}},
\bauthor{\bsnm{Feng}, \binits{Y.}},
\bauthor{\bsnm{Pan}, \binits{J.}},
\bauthor{\bsnm{Chen}, \binits{W.}}:
\batitle{G6: A web-based library for graph visualization}.
\bjtitle{Visual Informatics}
\bvolume{5}(\bissue{4}),
\bfpage{49}--\blpage{55}
(\byear{2021})
\end{barticle}
\endbibitem

\bibitem{franz2016cytoscape}
\begin{barticle}
\bauthor{\bsnm{Franz}, \binits{M.}},
\bauthor{\bsnm{Lopes}, \binits{C.T.}},
\bauthor{\bsnm{Huck}, \binits{G.}},
\bauthor{\bsnm{Dong}, \binits{Y.}},
\bauthor{\bsnm{Sumer}, \binits{O.}},
\bauthor{\bsnm{Bader}, \binits{G.D.}}:
\batitle{Cytoscape. js: a graph theory library for visualisation and analysis}.
\bjtitle{Bioinformatics}
\bvolume{32}(\bissue{2}),
\bfpage{309}--\blpage{311}
(\byear{2016})
\end{barticle}
\endbibitem

\bibitem{cherven2013network}
\begin{bbook}
\bauthor{\bsnm{Cherven}, \binits{K.}}:
\bbtitle{Network Graph Analysis and Visualization with Gephi}.
\bpublisher{Packt Publishing Ltd},
\blocation{Birmingham}
(\byear{2013})
\end{bbook}
\endbibitem

\bibitem{cheng2018building}
\begin{bchapter}
\bauthor{\bsnm{Cheng}, \binits{Y.}},
\bauthor{\bsnm{Liu}, \binits{F.C.}},
\bauthor{\bsnm{Jing}, \binits{S.}},
\bauthor{\bsnm{Xu}, \binits{W.}},
\bauthor{\bsnm{Chau}, \binits{D.H.}}:
\bctitle{Building big data processing and visualization pipeline through apache
  zeppelin}.
In: \bbtitle{Proceedings of the Practice and Experience on Advanced Research
  Computing},
pp. \bfpage{1}--\blpage{7}
(\byear{2018})
\end{bchapter}
\endbibitem

\bibitem{Wei2014}
\begin{barticle}
\bauthor{\bsnm{Wei}, \binits{H.}},
\bauthor{\bsnm{Yu}, \binits{J.X.}},
\bauthor{\bsnm{Lu}, \binits{C.}},
\bauthor{\bsnm{Jin}, \binits{R.}}:
\batitle{Reachability querying: An independent permutation labeling approach}.
\bjtitle{Proc. VLDB Endow.}
\bvolume{7}(\bissue{12}),
\bfpage{1191}--\blpage{1202}
(\byear{2014}).
\doiurl{10.14778/2732977.2732992}
\end{barticle}
\endbibitem

\bibitem{jin2019shortest}
\begin{botherref}
\oauthor{\bsnm{Jin}, \binits{R.}},
\oauthor{\bsnm{Ruan}, \binits{N.}}:
Shortest path computation in large networks.
Google Patents.
US Patent 10,521,473
(2019)
\end{botherref}
\endbibitem

\bibitem{mhedhbi2021a+}
\begin{bchapter}
\bauthor{\bsnm{Mhedhbi}, \binits{A.}},
\bauthor{\bsnm{Gupta}, \binits{P.}},
\bauthor{\bsnm{Khaliq}, \binits{S.}},
\bauthor{\bsnm{Salihoglu}, \binits{S.}}:
\bctitle{A+ indexes: Tunable and space-efficient adjacency lists in graph
  database management systems}.
In: \bbtitle{2021 IEEE 37th International Conference on Data Engineering
  (ICDE)},
pp. \bfpage{1464}--\blpage{1475}
(\byear{2021}).
\bcomment{IEEE}
\end{bchapter}
\endbibitem

\bibitem{10.1145/3282373.3282374}
\begin{bchapter}
\bauthor{\bsnm{Pokorn\'{y}}, \binits{J.}},
\bauthor{\bsnm{Valenta}, \binits{M.}},
\bauthor{\bsnm{Ramba}, \binits{J.}}:
\bctitle{Graph patterns indexes: Their storage and retrieval}.
In: \bbtitle{Proceedings of the 20th International Conference on Information
  Integration and Web-Based Applications Services}.
\bsertitle{iiWAS2018},
pp. \bfpage{221}--\blpage{225}.
\bpublisher{Association for Computing Machinery},
\blocation{New York, NY, USA}
(\byear{2018}).
\doiurl{10.1145/3282373.3282374}
\end{bchapter}
\endbibitem

\bibitem{sumrall2016investigations}
\begin{bchapter}
\bauthor{\bsnm{Sumrall}, \binits{J.M.}},
\bauthor{\bsnm{Fletcher}, \binits{G.H.}},
\bauthor{\bsnm{Poulovassilis}, \binits{A.}},
\bauthor{\bsnm{Svensson}, \binits{J.}},
\bauthor{\bsnm{Vejlstrup}, \binits{M.}},
\bauthor{\bsnm{Vest}, \binits{C.}},
\bauthor{\bsnm{Webber}, \binits{J.}}:
\bctitle{Investigations on path indexing for graph databases}.
In: \bbtitle{European Conference on Parallel Processing},
pp. \bfpage{532}--\blpage{544}
(\byear{2016}).
\bcomment{Springer}
\end{bchapter}
\endbibitem

\bibitem{sakr2012overview}
\begin{botherref}
\oauthor{\bsnm{Sakr}, \binits{S.}},
\oauthor{\bsnm{Al-Naymat}, \binits{G.}}:
An overview of graph indexing and querying techniques.
Graph Data Management: Techniques and Applications,
71--88
(2012)
\end{botherref}
\endbibitem

\bibitem{williams2007graph}
\begin{bchapter}
\bauthor{\bsnm{Williams}, \binits{D.W.}},
\bauthor{\bsnm{Huan}, \binits{J.}},
\bauthor{\bsnm{Wang}, \binits{W.}}:
\bctitle{Graph database indexing using structured graph decomposition}.
In: \bbtitle{2007 IEEE 23rd International Conference on Data Engineering},
pp. \bfpage{976}--\blpage{985}
(\byear{2007}).
\bcomment{IEEE}
\end{bchapter}
\endbibitem

\bibitem{besta2019demystifying}
\begin{botherref}
\oauthor{\bsnm{Besta}, \binits{M.}},
\oauthor{\bsnm{Peter}, \binits{E.}},
\oauthor{\bsnm{Gerstenberger}, \binits{R.}},
\oauthor{\bsnm{Fischer}, \binits{M.}},
\oauthor{\bsnm{Podstawski}, \binits{M.}},
\oauthor{\bsnm{Barthels}, \binits{C.}},
\oauthor{\bsnm{Alonso}, \binits{G.}},
\oauthor{\bsnm{Hoefler}, \binits{T.}}:
Demystifying graph databases: Analysis and taxonomy of data organization,
  system designs, and graph queries.
arXiv preprint arXiv:1910.09017
(2019)
\end{botherref}
\endbibitem

\bibitem{patil2018survey}
\begin{barticle}
\bauthor{\bsnm{Patil}, \binits{N.}},
\bauthor{\bsnm{Kiran}, \binits{P.}},
\bauthor{\bsnm{Kiran}, \binits{N.}},
\bauthor{\bsnm{KM}, \binits{N.P.}}:
\batitle{A survey on graph database management techniques for huge unstructured
  data}.
\bjtitle{International Journal of Electrical and Computer Engineering}
\bvolume{8}(\bissue{2}),
\bfpage{1140}
(\byear{2018})
\end{barticle}
\endbibitem

\bibitem{besta2018survey}
\begin{botherref}
\oauthor{\bsnm{Besta}, \binits{M.}},
\oauthor{\bsnm{Hoefler}, \binits{T.}}:
Survey and taxonomy of lossless graph compression and space-efficient graph
  representations.
arXiv preprint arXiv:1806.01799
(2018)
\end{botherref}
\endbibitem

\bibitem{Angles2018}
\begin{bbook}
\bauthor{\bsnm{Angles}, \binits{R.}},
\bauthor{\bsnm{Gutierrez}, \binits{C.}}:
In: \beditor{\bsnm{Fletcher}, \binits{G.}},
\beditor{\bsnm{Hidders}, \binits{J.}},
\beditor{\bsnm{Larriba-Pey}, \binits{J.L.}} (eds.)
\bbtitle{An Introduction to Graph Data Management},
pp. \bfpage{1}--\blpage{32}.
\bpublisher{Springer},
\blocation{Cham}
(\byear{2018}).
\doiurl{10.1007/978-3-319-96193-4_1}
\end{bbook}
\endbibitem

\bibitem{Sakr2021-dl}
\begin{barticle}
\bauthor{\bsnm{Sakr}, \binits{S.}},
\bauthor{\bsnm{Bonifati}, \binits{A.}},
\bauthor{\bsnm{Voigt}, \binits{H.}},
\bauthor{\bsnm{Iosup}, \binits{A.}},
\bauthor{\bsnm{Ammar}, \binits{K.}},
\bauthor{\bsnm{Angles}, \binits{R.}},
\bauthor{\bsnm{{Yoneki}}}:
\batitle{The future is big graphs: a community view on graph processing
  systems}.
\bjtitle{Communications of the ACM}
\bvolume{64}(\bissue{9}),
\bfpage{62}--\blpage{71}
(\byear{2021})
\end{barticle}
\endbibitem

\bibitem{Sahu2020-pi}
\begin{barticle}
\bauthor{\bsnm{Sahu}, \binits{S.}},
\bauthor{\bsnm{Mhedhbi}, \binits{A.}},
\bauthor{\bsnm{Salihoglu}, \binits{S.}},
\bauthor{\bsnm{Lin}, \binits{J.}},
\bauthor{\bsnm{{\"O}zsu}, \binits{M.T.}}:
\batitle{The ubiquity of large graphs and surprising challenges of graph
  processing: extended survey}.
\bjtitle{VLDB J.}
\bvolume{29}(\bissue{2-3}),
\bfpage{595}--\blpage{618}
(\byear{2020})
\end{barticle}
\endbibitem

\bibitem{fernandes2018graph}
\begin{bchapter}
\bauthor{\bsnm{Fernandes}, \binits{D.}},
\bauthor{\bsnm{Bernardino}, \binits{J.}}:
\bctitle{Graph databases comparison: Allegrograph, arangodb, infinitegraph,
  neo4j, and orientdb.}
In: \bbtitle{Data},
pp. \bfpage{373}--\blpage{380}
(\byear{2018})
\end{bchapter}
\endbibitem

\bibitem{meiling2017benchmarking}
\begin{botherref}
\oauthor{\bsnm{Meiling}, \binits{L.}}, et al.:
Benchmarking multi-model databases with arangodb and orientdb
(2017)
\end{botherref}
\endbibitem

\bibitem{Macrobenchmarks}
\begin{barticle}
\bauthor{\bsnm{Lissandrini}, \binits{M.}},
\bauthor{\bsnm{Brugnara}, \binits{M.}},
\bauthor{\bsnm{Velegrakis}, \binits{Y.}}:
\batitle{Beyond macrobenchmarks: Microbenchmark-based graph database
  evaluation}.
\bjtitle{Proc. VLDB Endow.}
\bvolume{12}(\bissue{4}),
\bfpage{390}--\blpage{403}
(\byear{2018}).
\doiurl{10.14778/3297753.3297759}
\end{barticle}
\endbibitem

\bibitem{LDBC2020}
\begin{botherref}
\oauthor{\bsnm{Angles}, \binits{R.}},
\oauthor{\bsnm{Antal}, \binits{J.B.}},
\oauthor{\bsnm{Averbuch}, \binits{A.}},
\oauthor{\bsnm{Boncz}, \binits{P.}},
\oauthor{\bsnm{Erling}, \binits{O.}},
\oauthor{\bsnm{Gubichev}, \binits{A.}},
\oauthor{\bsnm{Haprian}, \binits{V.}},
\oauthor{\bsnm{Kaufmann}, \binits{M.}},
\oauthor{\bsnm{Pey}, \binits{J.L.L.}},
\oauthor{\bsnm{Martínez}, \binits{N.}},
\oauthor{\bsnm{Marton}, \binits{J.}},
\oauthor{\bsnm{Paradies}, \binits{M.}},
\oauthor{\bsnm{Pham}, \binits{M.-D.}},
\oauthor{\bsnm{Prat-Pérez}, \binits{A.}},
\oauthor{\bsnm{Spasić}, \binits{M.}},
\oauthor{\bsnm{Steer}, \binits{B.A.}},
\oauthor{\bsnm{Szárnyas}, \binits{G.}},
\oauthor{\bsnm{Waudby}, \binits{J.}}:
The LDBC Social Network Benchmark.
arXiv
(2020).
\doiurl{10.48550/ARXIV.2001.02299}
\end{botherref}
\endbibitem

\bibitem{mhedhbi2021lsqb}
\begin{bchapter}
\bauthor{\bsnm{Mhedhbi}, \binits{A.}},
\bauthor{\bsnm{Lissandrini}, \binits{M.}},
\bauthor{\bsnm{Kuiper}, \binits{L.}},
\bauthor{\bsnm{Waudby}, \binits{J.}},
\bauthor{\bsnm{Sz{\'a}rnyas}, \binits{G.}}:
\bctitle{Lsqb: a large-scale subgraph query benchmark}.
In: \bbtitle{Proceedings of the 4th ACM SIGMOD Joint International Workshop on
  Graph Data Management Experiences \& Systems (GRADES) and Network Data
  Analytics (NDA)},
pp. \bfpage{1}--\blpage{11}
(\byear{2021})
\end{bchapter}
\endbibitem

\bibitem{rusu2019depth}
\begin{botherref}
\oauthor{\bsnm{Rusu}, \binits{F.}},
\oauthor{\bsnm{Huang}, \binits{Z.}}:
In-depth benchmarking of graph database systems with the linked data benchmark
  council (ldbc) social network benchmark (snb).
arXiv preprint arXiv:1907.07405
(2019)
\end{botherref}
\endbibitem

\bibitem{wangproc2020empriical}
\begin{bchapter}
\bauthor{\bsnm{Wang}, \binits{R.}},
\bauthor{\bsnm{Yang}, \binits{Z.}},
\bauthor{\bsnm{Zhang}, \binits{W.}},
\bauthor{\bsnm{Lin}, \binits{X.}}:
\bctitle{An empirical study on recent graph database systems}.
In: \beditor{\bsnm{Li}, \binits{G.}},
\beditor{\bsnm{Shen}, \binits{H.T.}},
\beditor{\bsnm{Yuan}, \binits{Y.}},
\beditor{\bsnm{Wang}, \binits{X.}},
\beditor{\bsnm{Liu}, \binits{H.}},
\beditor{\bsnm{Zhao}, \binits{X.}} (eds.)
\bbtitle{Knowledge Science, Engineering and Management},
pp. \bfpage{328}--\blpage{340}.
\bpublisher{Springer},
\blocation{Cham}
(\byear{2020})
\end{bchapter}
\endbibitem

\end{thebibliography}


\section{Appendices}

\end{document}